\documentclass[12pt, a4paper]{article}
\title{Quasicrystals. The Technique for Constructing Quasiperiodic Structures}
\author{V.~C.~Gulyaev\footnote{e-mail: gulk\_r@yahoo.com}\\
\itshape Krasnoyarsk, Russia}
\date{March 11, 1999}

\usepackage{amssymb, amsmath}
\usepackage[dvips]{graphicx,color}
\usepackage{calc}
\newcommand{\captn}[1]{\caption{\footnotesize #1}}
\begin{document}
\maketitle
\begin{abstract}
In this paper, a technique for constructing quasiperiodic structures
is suggested, which allows one by the assigned matching to restore
the atoms density distribution formula of a corresponding
quasicrystal. The algorithm to restore the atom density
distribution has been considered on the example of the Penrose
matching. The analytical record of a Penrose quasicrystal is
given\footnote{See program in files pict19.ps--pict21.ps too.}.
\end{abstract}
\section       {Introduction} \label{cint}
In the basis of traditional crystallography is the conception of periodically
repeated structures. I.e. in crystals there is realized such a packing of one
structural cell, which possesses translational symmetry in respect to a shift
by some vector. Translational symmetry is a definition of a long-range order
in crystals. Simultaneously with translational symmetry, rotational symmetry
through angles
\begin{equation}
\label{math/1}
\qquad \phi=\frac{2\pi}{k} \qquad (k=1,2,3,4,6)
\end{equation}
can also exist.
The same situation was in the studies on tilings, packings, mosaics etc.:
it was believed that if there existed not one, but two or several matching
elements, one could find a method of periodic tiling of a plane with the
elements. Such tilings also possess rotational symmetries (\ref{math/1}).
However, in 1974 R. Penrose [1] showed that with several elements, two as
minimum, one could match a plane quasiperiodically. Such quasiperiodic
matchings can have symmetry axes, different from symmetry exes allowed for
periodic structures. (The concept
of symmetry for quasiperiodic structures is still to be defined). Penrose's
mathematic discovery anticipated the existence of quasiperiodic structures
obtained experimentally in metal alloys. In 1984 an article was published [2],
in which an experimental evidence was presented for the existence of an alloy
with exceptional properties. An X-ray photograph showed a sharp system of spots
testifying for the existence of a pentagonal order (to be more exact,
decagonal), and the availability of a long-range order. At present, a great
number of papers have been published devoted both to experimental and
theoretical studies of such structures called quasicrystals. There exist several
methods to describe quasicrystals: a technique based on projecting on a 2D or 3D
space of a periodic hyperlattice of a higher dimensionality [3-6], another
method of projecting [7] consists in constructing periodic packings of
icosahedrons or pentagons in curved 3D and 2D spaces respectively, where
symmetry exclusions are weaker than in zero curviture spaces. But a return to
the real Euclide world requires to introduce disclination lines which generate
a long-range icosahedral order for a 3D space or a long-rang order
corresponding to the Penrose matching for a 2D space. In the works [8-11] a
detailed analysis of the Penrose problem has been undertaken, the GDM
(generalized dual method) rules formulated [8,9] or multigrid ones [10,11],
allowing one to construct quasicrystals of an infinite extent with icosahedral
symmetry (3D) and point symmetry of the fifth order (2D). A cluster approach
to consider the quasicrystal problem has been suggested in [12].
In the work, a technique to describe quasicrystal structures is suggested,
using  a traditional approach to describe ideal (periodic)
crystals and incommensurate structures. In particular the symmetries of
quasicrystal structures are considered with the superspace groups [13,14],
which now are a classical method of describing incommensurate structures.
\section                    {1D quasicrystals}  \label{c1dqc}
\begin{figure}[pht]
\centering
\includegraphics{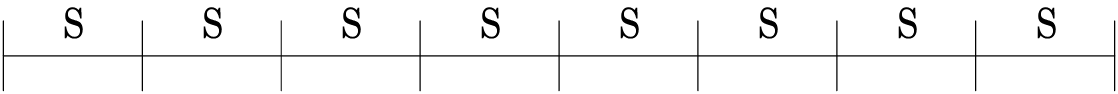}
\captn{The periodic sequence of one interval of the length $\mathrm{S}=1.$}
\label{f:1}
\end{figure}

\begin{figure}[h]
\centering
\includegraphics{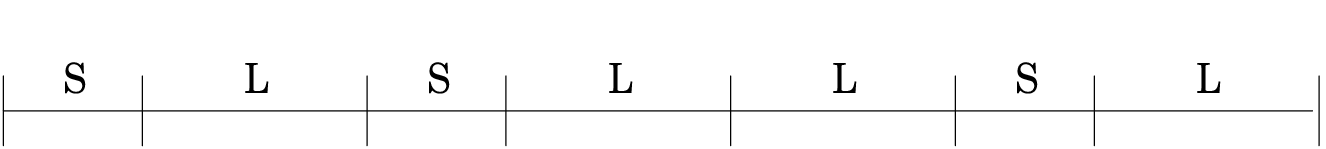}
\captn{The quasiperiodic sequence of two intervals L and S.} \label{f:2}
\end{figure}

\begin{figure}[hpt]
\centering
\includegraphics{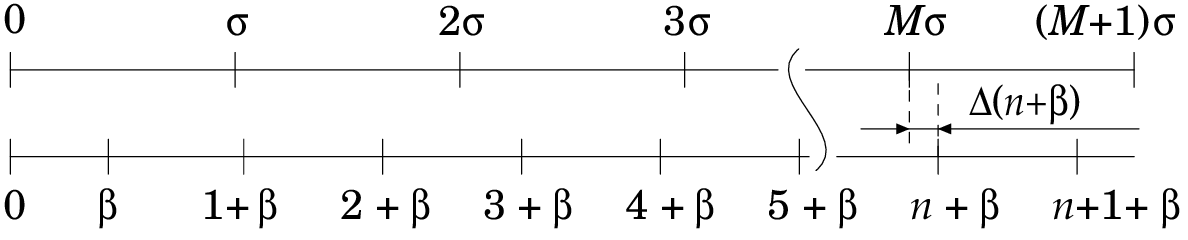}
\captn{$\Delta(n+\beta)=n+\beta-{\sigma}M=
\sigma \left\{(n+\beta)/\sigma \right\}$.}
\label{f:4}
\end{figure}

\begin{figure}[hpt]
\centering
\includegraphics{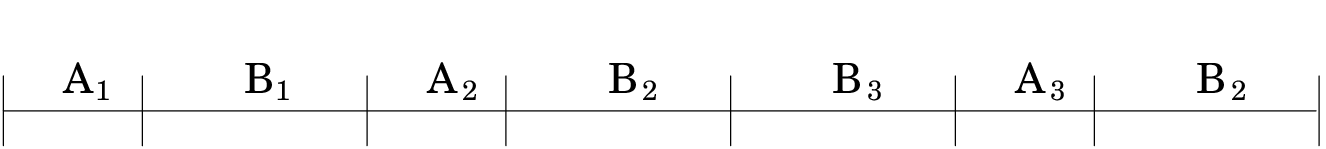}
\captn{The quasiperiodic sequence of six structural cells with
the lengths L and S.} \label{f:5}
\end{figure}

Even though there is no analogue of orientational order in one dimension,
one-dimensional quasicrystals will be useful to illustrate some important
properties of quasiperiodic structures. To begin with, remember the principle
of constructing periodic crystals. Consider a one-dimensional periodic lattice
with the position of sites at the points
\begin{equation}
\label{math/2}
x_n=n.
\end{equation}
In this case we have a periodic sequence of one interval of the length
$\mathrm{S}=1$, as shown in Fig.\ref{f:1}.
To each point of the lattice we relate some group of atoms---the basis, with
all the groups being identical by the composition and arrangenment. The basis,
consisting of $J$ atoms, is defined by a $J$ set of $x_j$ coordinates which
determine the positions of the atom centres with respect to a lattice
point, to which the basis is related. Atoms constituting the basis are arranged
with respect to a given lattice point in such a way that
\mbox{$0 \leqslant x_j< \mathrm{S}$}. A set of atoms, thus assigned, is a
unit cell of a periodic structure. For such a structure, the atoms density
distribution (we will consider atoms as points) $\rho(x)$ can be written in
the form of the sum
\begin{equation}
\label{math/3}
\rho\,(x)=\sum_{n=-\infty}^{\infty}\sum_{j=1}^{J}\delta\left(x-n-x_j\right),
\end{equation}
where the first summation $(j=1,...,J)$ is taken with respect to all basis
atoms, and the second -- to all lattice sites. Quasiperiodic structures, in
contrast to periodic structures, consist of several structural cells forming a
quasiperiodic sequence. Structural cells of a quasicrystal can differ both by
the number of atoms and their arrangement and by sizes. Consider a quasicrystal
consisting of two structural cells of the same size, but with different bases.
Denote the internal atom coordinates of the first and second bases of
$x_{j_1}$, $x_{j_2}$ respectively. In order to obtain the atoms density
distribution formula for a quasicrystal, analogous to (\ref{math/3}), introduce
the function
\begin{equation}
\label{math/4}
p(n+\beta)=\left\lfloor\frac{n+1+\beta}{\sigma}\right\rfloor-
\left\lfloor\frac{n+\beta}{\sigma}\right\rfloor,
\end{equation}
where $\lfloor{x}\rfloor$ is an greatest integer part of the number, $\sigma>1$
is an irrational number, $\beta$ is the arbitrary number.
The function $p(n+\beta)\ (n=0,\pm{1},\pm{2}\ldots)$ assigns a quasiperiodic
sequence of zeros and units [10,15-17]. In the sequence the relation of the
number of units to the number of zeros equals $1/(\sigma-1)$. Define the
functions
\begin{equation}
\label{math/5}
\begin{array}{l}
p_1(n+\beta)=1-{p}(n+\beta)\, , \\
p_2(n+\beta)={p}(n+\beta)
\end{array}
\end{equation}
as probabilities of arranging the first and the second bases in a cell at the
assigned $n$. For such a structure the atoms density distribution
$\rho(x,\beta)$ can be written in the form of the sum
\begin{equation}
\label{math/6}
\rho(x,\beta)=\sum_{n=-\infty}^{\infty}\sum_{k=1}^{2}p_{k}(n+\beta)
\sum_{j_k=1}^{J_k}\delta\left(x-n-x_{j_k}\right),
\end{equation}
where $0\leqslant{x_j}_k<\mathrm{S}$, the first summation $(j_k=1,..,J_k)$ is
made in all atoms of the basis, the second summation $(k=1,2)$-- in all bases,
and the third-- in all sites of the lattice. In the case of a quasiperiodic
structure with two intervals $\mathrm{L}$ and $\mathrm{S}$ between the
neighboring sites, position of sites at the $x_n$ points is assigned by
the formula
\begin{equation}
\label{math/7}
x_{n}(\alpha,\beta)=n+\alpha+\frac{1}{\chi}
\left\lfloor\frac{n+\beta}{\sigma} \right\rfloor,
\end{equation}
where $\chi>0$, $\alpha$ is an arbitrary number, further we will consider
\mbox{$\beta\left/\sigma\right.<1$}, since in the opposite case $\alpha$
changes, i.e. $\alpha'=\alpha+1/\chi{\left\lfloor{\beta/\sigma}\right\rfloor}$.
The relationship (\ref{math/7}) describes the position of the lattice sites so
that the interval \mbox{$\Delta{x_n}=x_{n+1}-x_{n}$} between the sites has the
property:
\begin{equation}
\label{math/8}
\Delta{x_n}=\left|
\begin{array}{ll}
\mathrm{S}=1\:, & \mbox{if } \; p_1(n+\beta)=1\:, \\ \displaystyle
\mathrm{L}=1+\frac{1}{\chi}\:,  & \mbox{if } \; p_2(n+\beta)=1 \:.
\end{array}\right.
\end{equation}
The parameter $\chi$ determines the relationship between the interval lengths
in the sequence. For instance, for a particular case of
$\sigma=\chi=\tau=(1+\sqrt5\,)/2$ the quasiperiodic sequence of two intervals
is shown in Fig.\ref{f:2}.
In this sequence the interval  L is repeated once or two times in succession,
the interval S only once. For a quasiperiodic structures with two structural
cells of different length, the atoms density distribution
$\rho(x,\alpha,\beta)$ can be written in the form of the sum
\begin{equation}
\label{math/9}
\rho(x,\alpha,\beta)=\sum_{n=-\infty}^{\infty}\sum_{k=1}^{2}{p_{k}(n+\beta)}
\sum_{j_k=1}^{J_k}\delta\left( x-x_{n}(\alpha,\beta)-x_{j_k} \right),
\end{equation}
where $0 \leqslant x_{j_1}<\mathrm{S}$, $0 \leqslant x_{j_2}<\mathrm{L}$.
To further study structural properties of quasicrystals, make a detailed
analysis of the functions $p_{k}(n+\beta)$. To demonstrate, consider the function of
a continuous variable $x$
\begin{equation}
\label{math/10}
f(x,{\delta}_1,{\delta}_2)=\left\lfloor  \frac{x+\delta_1}{\sigma} \right\rfloor
-\left\lfloor{\frac{x+\delta_2}{\sigma}}\right\rfloor,
\end{equation}
where ${\sigma}\geqslant{\delta_1}>{\delta_2}$. The function
$f(x,{\delta_1},{\delta_2})$ is periodic in $x$ with the period $\sigma$. The
graph $f(x,{\delta_1},{\delta_2})$ represents a sequence of impulses with the
length of $(\delta_1-\delta_2)$ and the distance between the impulses
\begin{figure}[hbt]
\centering
\includegraphics{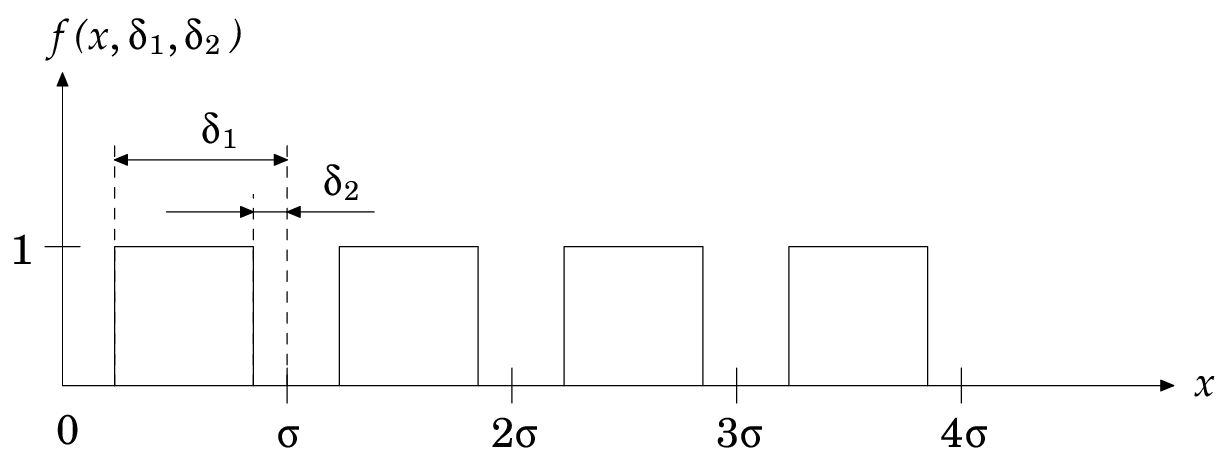}
\captn{The graph $ f(x,{\delta_1},{\delta_2})$ represents a sequence
of impulses with the length of $(\delta_1-\delta_2)$ and the distance between
the impulses $({\sigma}-{{\delta}_1}+{{\delta}_2}).$} \label{f:3}
\end{figure}
$({\sigma}-{{\delta}_1}+{{\delta}_2})$ (Fig.\ref{f:3}).
For $f(x,{\delta_1},{\delta_2})$ the following equation holds true:
\[f(x+\varepsilon,{\delta_1},{\delta_2})=f(x,{\delta_1}+
\varepsilon,{\delta_2}+\varepsilon).\]
Moving along the $x$ axis of the graph of the function $f(x, \delta_1, \delta_2)$
to $n$ from the initial position of $x=\beta$ will
give the value of the function
\begin{equation}
\label{math/11}
f(n+\beta,{\delta}_1,{\delta}_2)=\left\lfloor{\frac{n+\beta+\delta_1}{\sigma}}
\right \rfloor-\left\lfloor{\frac{n+\beta+{\delta}_2}{\sigma}}\right\rfloor.
\end{equation}
For example: $f(n+\beta,\sigma,1)=p_1(n+\beta)$, $f(n+\beta,1,0)=p_2(n+\beta)$.
Calculate the position of the point $\Delta(n+\beta)$ within the period $\sigma$ when
shifting from the origin of coordinates by the value of $n+\beta$.
To do this, calculate first by what maximum integer $M$ of $\sigma$ periods the
movement occurred:
\begin{equation}
\label{math/12}
M=\left\lfloor \frac{n+\beta}{\sigma} \right\rfloor \,.
\end{equation}
Then (Fig.\ref{f:4})
\begin{equation}
\label{math/13}
\Delta(n+\beta)=n+\beta-{\sigma}M=
\sigma \left\{ \frac{n+\beta}{\sigma} \right\},
\end{equation}
where $\{x\}$ is a fractional part of the number, $\{x\}=x-\lfloor x \rfloor$.
Taking account of the periodicity $\Delta(n+\beta)$, the function
$f(n+\beta,\delta_1,\delta_2)$ can be defined as follows:
\begin{equation}
\label{math/14}
f(n+\beta,\delta_1,\delta_2)=
\left|
\begin{array}{ll}
0, & \mbox{if } \; 0 \leqslant \Delta (n+\beta) < \sigma-{\delta}_1,\:
               \sigma-\delta_2 \leqslant \Delta(n+\beta) < \sigma, \\
1, & \mbox{if } \; \sigma - \delta_1 \leqslant \Delta(n+\beta)< \sigma-\delta_2 \,.
\end{array}
\right.
\end{equation}
Such a definition of $f(n+\beta,\delta_1,\delta_2)$, equivalent to the
definition of this function in (\ref{math/11}), will be useful in studying 2D and
3D quasicrystals. Divide the period $\sigma$ into several sections of the length
\begin{equation}
\label{math/15}
s_1= \sigma - \delta_1, \; s_2=\delta_1-\delta_2, \ldots,\;
s_{K-1}= \delta_{K-2}- \delta_{K-1}, \; s_K=\delta_{K-1},
\end{equation}
where $\sigma > \delta_1 > \delta_2 > \ldots > \delta_{K-1}$.
Now the number of structural cells of the same size, but with different bases,
can be increased until $K$, defining $K$ probabilities as follows:
\begin{equation}
\label{math/16}
\begin{array}{l}
p_1(n+\beta)=f(n+\beta,\sigma,\delta_1),\;
p_2(n+\beta)=f(n+\beta,\delta_1,\delta_2), \ldots, \\
p_{K-1}(n+\beta)=f(n+\beta,\delta_{K-2},\delta_{K-1}),\\
p_{K}(n+\beta)=f(n+\beta,\delta_{K-1},0).
\end{array}
\end{equation}
For such a structure, the atoms density distribution $\rho(x,\beta)$ can be
written in the form of the sum
\begin{equation}
\label{math/17}
\rho(x,\beta)=\sum_{n=-\infty}^{\infty} \sum_{k=1}^{K} p_{k}(n+\beta)
\sum_{j_k=1}^{J_k} \delta\left(x-n-x_{j_k}\right),
\end{equation}
where $0 \leqslant x_{j_k} < \mathrm{S} \; \;(k=1,2,\ldots,K)$. Repetition
frequency of the basis with $k$ index in such a structure will be equal to
$s_k/\sigma$. In the case of a quasiperiodic structure with two intervals
defined in formula (\ref{math/8}), divide the period $\sigma$ into sections
as follows:
\begin{equation}
\label{math/18}
s_1=\sigma-\delta_1,\;s_2=\delta_1-\delta_2,\;\ldots,\;s_{m}=\delta_{m-1}-1\,,
\end{equation}
where $\sigma>{\delta}_1 > \delta_2 > \ldots> \delta_{m-1} > 1$.
\begin{equation}
\label{math/19}
\begin{array}{l}
s_{m+1}=1-\delta_m,\; s_{m+2}=\delta_m-\delta_{m+1},\ldots, \\
s_{K-1}=\delta_{K-2}-\delta_{K-1}, \; s_{K}=\delta_{K-1},
\end{array}
\end{equation}
where $ 1 > \delta_m > \delta_{m+1} > \ldots > \delta_{K-1} > 0$.
Then the first $m$ of the probabilities $p_1(n+\beta),..p_m(n+\beta)$, defined
in (\ref{math/11}), will correspond to the arrangement of $m$ bases within the
interval with the length $S$, and $(K-m)$ of the probabilities
$p_{m+1}(n+\beta), \ldots ,p_K(n+\beta)$ --- to the arrangement of $(K-m)$ bases
within the interval with the length L. For such a quasiperiodic structure,
the atoms density distribution $p(x,\beta)$ can be written in the form of the
sum
\begin{equation}
\label{math/20}
\rho(x,\beta)=\sum_{n=-\infty}^{\infty}\sum_{k =1}^{K}{p_{k}(n+\beta)}
\sum_{j_k=1}^{J_k}{\delta\left(x-x_n(\alpha,\beta)-x_{j_k}\right)},
\end{equation}
where $0 \leqslant x_{j_k} < \mathrm{S} \;(k=1,2,\ldots,m), \;
0 \leqslant x_{j_k} < \mathrm{L} \; (k=m+1,m+2,\ldots,K)$.
To construct a quasiperiodic structure of an assigned configuration with bases
alternating in a certain manner, take advantage of a famous fact from the
probability theory. If $p_1, p_2,\ldots,\, p_m$ are the probabilities of setting
in $m$ independent developments, then
\begin{equation}
\label{math/21}
p=\prod_{k =1}^{m} p_k
\end{equation}
is the probability of setting in all $m$ developments simultaneously.
To illustrate the above, consider a particular example: $\chi=\sigma=\tau$, and
$p_1, p_2, x_n$ are taken from formulae (\ref{math/5}), (\ref{math/7}) respectively. The sequence of
intervals for $x_n(\alpha,\beta)$ will correspond to Fig.\ref{f:2}. For the given
quasiperiodic sequence of the intervals, introduce some structural cells with
the lengths L and S with different bases (Fig.\ref{f:5}).
As seen from Fig.\ref{f:5}, structural cells with the length S are arranged as
follows:
on the right of the cell $A_1$, in one interval, is a cell with the length S;
on the left of the cell $A_2$, in one interval, is a cell with the length S;
on both sides of the cell $A_3$, in one interval, is a cell with the length L.
Structural cells with the length L in a sequence are arranged as follows:
$B_1$ -- on both sides there are two cells with the length S; $B_2$, $B_3$
-- are behind each other: $B_2$ is the left, $B_3$ the right structural cells.
According to the above said, define probabilities corresponding to these
structural cells
\begin{subequations}
\label{math/22}
\allowdisplaybreaks
\begin{align}
{p_A}_{1}(n+\beta)&=p_{1}(n+\beta)p_{1}(n+2+\beta) \,,\\
{p_A}_{2}(n+\beta)&=p_{1}(n+\beta)p_{1}(n-2+\beta) \,,\\
{p_A}_{3}(n+\beta)&=p_{1}(n+\beta)p_{2}(n-2+\beta)p_{2}(n+2+\beta) \,,\\
{p_B}_{1}(n+\beta)&=p_{2}(n+\beta)p_{1}(n-1+\beta)p_{1}(n+1+\beta) \,,\\
{p_B}_{2}(n+\beta)&=p_{2}(n+\beta)p_{2}(n+1+\beta)\,,\\
{p_B}_{3}(n+\beta)&=p_{2}(n+\beta)p_{2}(n-1+\beta)\,.
\end{align}
\end{subequations}
Calculations for equations (\ref{math/22}) (refer to Appendix \ref{cfpr})
produce the following result.
\begin{subequations}
\label{math/23}
\allowdisplaybreaks
\begin{align}
p_{A_1}(n+\beta) &=\left\lfloor \frac{n+\beta+\tau}{\tau} \right\rfloor -
\left\lfloor \frac{n+1+1/\tau^2+\beta}{\tau} \right\rfloor
\vphantom{ \cfrac{1}{ 1+\frac{1}{ 1+\frac{1}{1}}}} \,,\\
p_{A_2}(n+\beta) &= \left\lfloor \frac{n+\beta+2/\tau}{\tau} \right\rfloor -
\left\lfloor \frac{n+\beta+1}{\tau} \right\rfloor
\vphantom{ \cfrac{1}{ 1+\frac{1}{ 1+\frac{1}{1} } } } \,,\\
p_{A_3}(n+\beta) &= \left\lfloor \frac{n+\beta+1+ 1 /\tau^2}{\tau}
\right\rfloor - \left\lfloor \frac{n+\beta+2/\tau}{\tau} \right\rfloor
\vphantom{ \cfrac{1}{ 1+\frac{1}{ 1+\frac{1}{1} } } }  \,,\\
p_{B_1}(n+\beta) &= \left\lfloor \frac{n+\beta+1/\tau}{\tau} \right\rfloor -
\left\lfloor \frac{n+\beta +1/\tau^2}{\tau} \right\rfloor
\vphantom{ \cfrac{1}{ 1+\frac{1}{ 1+\frac{1}{1} } } }  \,,\\
p_{B_2}(n+\beta) &= \left\lfloor \frac{n+\beta+1}{\tau} \right\rfloor -
\left\lfloor \frac{n+\beta+1/\tau}{\tau} \right\rfloor
\vphantom{ \cfrac{1}{ 1+\frac{1}{ 1+\frac{1}{1} } } }   \,,\\
p_{B_3}(n+\beta) &= \left\lfloor \frac{n+\beta+1/\tau^2}{\tau}\right
\rfloor - \left\lfloor \frac{n+\beta}{\tau} \right\rfloor
\vphantom{ (\cfrac{1}{ 1+\frac{1}{ 1+\frac{1}{1} } })^1 } \,.
\end{align}
\end{subequations}
These probabilities assign quasiperiodic sequences of zeros and units
$(n=0,1,\ldots,61; \beta=0)$ \\
${p_A}_1$: $10000100000001000010000000100000001000010000000100001000000010$
${p_A}_2$: $00000001000000010000100000001000000010000100000001000010000000$
${p_A}_3$: $00100000001000000000000100000001000000000000100000000000010000$
${p_B}_1$: $00000010000000100001000000010000000100001000000010000100000001$
${p_B}_2$: $01001000010010000100001001000010010000100001001000010000100100$
${p_B}_3$: $00010000100100001000010010000100100001000010010000100001001000$
%
\section                   {2D and 3D quasicrystals}  \label{c2d3d}
\begin{figure}[hpt]
\centering
\includegraphics{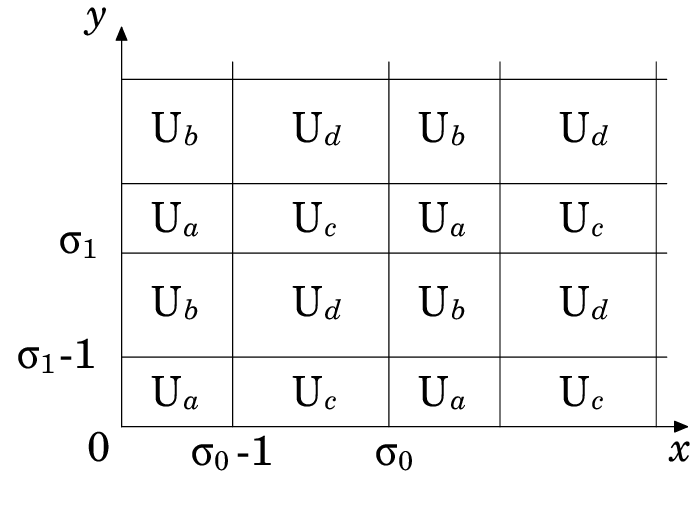}
\captn{The areas U$_a$, U$_b$, U$_c$, U$_d$, in which probabilities
$p_a$, $p_b$, $p_c$, $p_d$ assume values equal to unity respectively.}
\label{f:6}
\end{figure}

\begin{figure}[htb]
\includegraphics*[-1,-1][190,190]{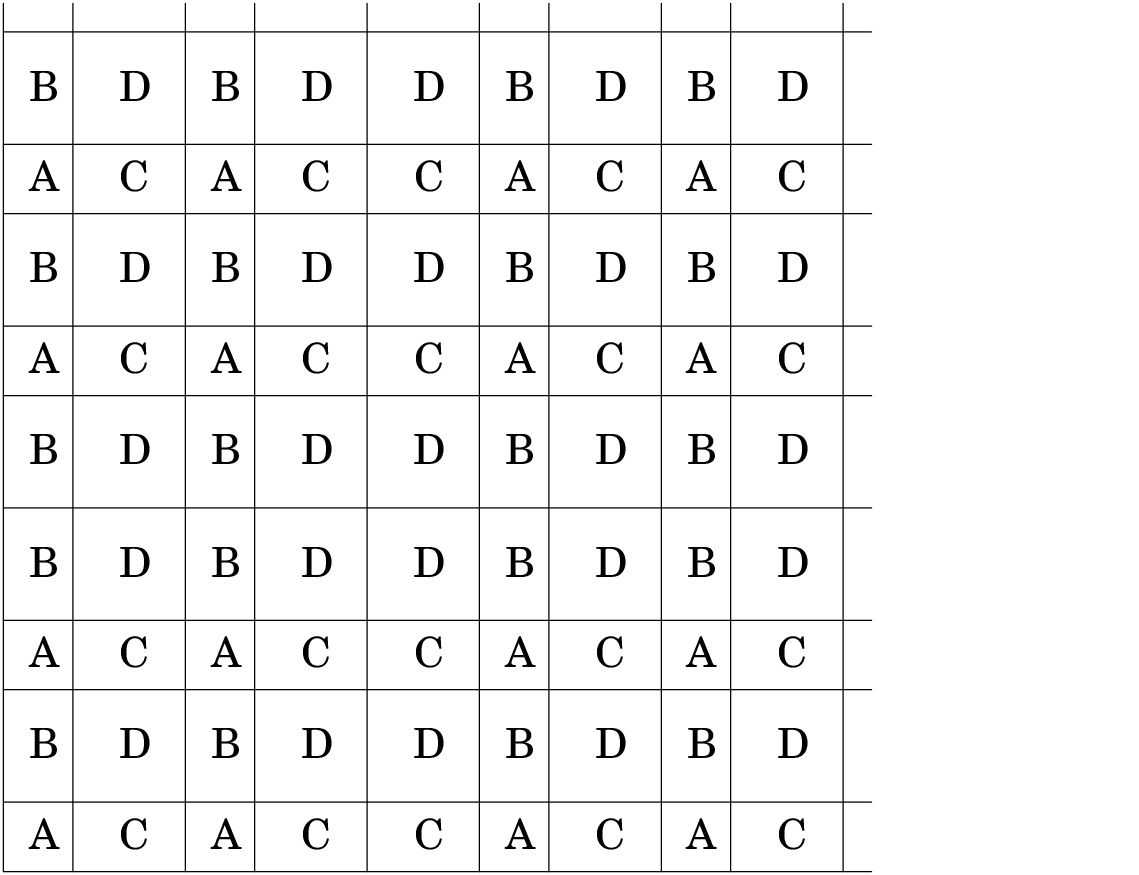}
\hfill
\includegraphics*{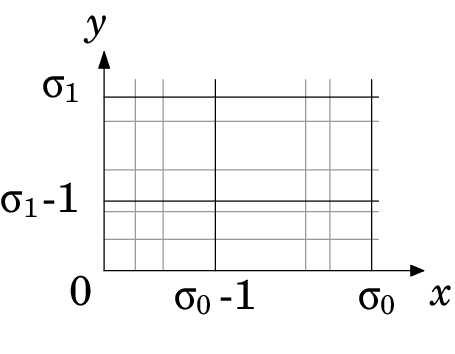}
\parbox[t]{190pt}{\captn{The quasicrystal, assigned by formulae
(\ref{math/28}) for a particular case of \mbox{$\sigma_0=\sigma_1=\tau$.}}
\label{f:7}}
\hfill
\parbox[t]{171pt}{\captn{In increasing the number of structural cells,
analogously to (\ref{math/16}), each region in the graph Fig.\ref{f:6}
will be divided into several rectangulars.}
\label{f:8}}
\end{figure}

\begin{figure}[hpt]
\centering
\includegraphics{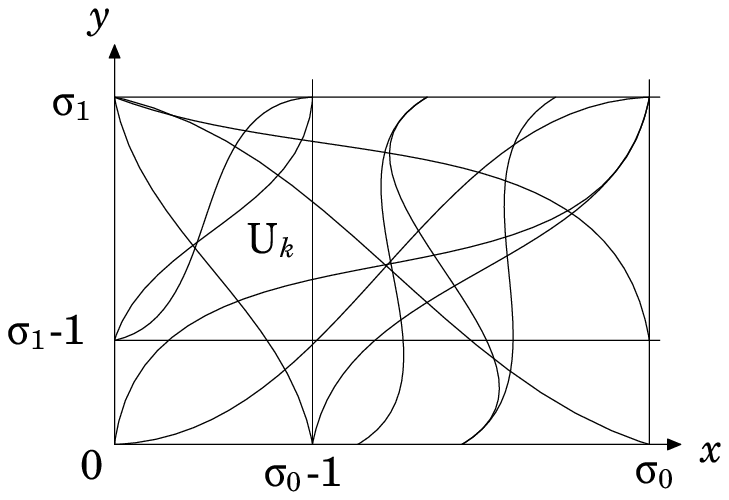}
\captn{In a general case, probabilities will be corresponded
by regions of an arbitrary form restrained by curves.}
\label{f:9}
\end{figure}

The orientational symmetry is a new element in extending a quasiperiodic
translational order from one  dimension to two or three ones. 2D and 3D
quasicrystals allow symmetries excluded for periodic crystals. A simple
generalization of one--dimensional quasicrystals will be 3D quasicrystals with
vectors of the main translations $\mathbf{a}_0$, $\mathbf{a}_1$,
$\mathbf{a}_2$ of the following form:\\
\begin{equation}
\label{math/24}
\begin{split}
\rho (\mathbf{r},\boldsymbol{\beta}) =
& \sum\limits^{\infty}_{{n_0},\, {n_1},\, {n_2} = -\infty \;}
\sum\limits_{{k_0},\,{k_1},\,k_2 = 1}^{{K_0},\,{K_1},\,K_2}
p_{k_0}(n_0+\beta_0) p_{k_1}(n_1+\beta_1)
{p_k}_3(n_2+\beta_2) \vphantom{ \cfrac{1}{ 1+\frac{1}{ 1+\frac{1}{1} } } } \\
& \quad \cdot \sum\limits_{j_{k_0k_1k_2}=1}^{J_{k_0k_1k_2}}
\delta \bigl( \mathbf{r-r_n}(\boldsymbol{\alpha},{\boldsymbol\beta}) -
\mathbf{r}_{j_{{\,k_0}{k_1}{k_2}}} \bigr),
 \end{split}
\end{equation}
where $\mathbf{r_n}(\boldsymbol{\alpha,\beta})=
\mathbf{a}_0 x_{n_0}+\mathbf{a}_1x_{n_1}+
\mathbf{a}_2x_{n_2}$,\\
${\mathbf{r}_j}_{k_0k_1k_2}=\mathbf{a}_0{x_j}_{k_0k_1k_2}+\mathbf{a}_1{y_j}_{k_0k_1k_2}+
\mathbf{a}_2{z_j}_{k_0k_1k_2}$ is the vector
determining the location of the basis atom centres with the index $k_0k_1k_2$
relative to the lattice point to which the basis is related,
$x_{n_l}=n_l+\alpha_l+(1/\chi_l)\left\lfloor(n_l+\beta_l)/
\sigma_l\right\rfloor$.
Atoms forming the basis are arranged relative to a given lattice site in such a
way that \; $0 \leqslant {x_j}_{k_0k_1k_2}<{\mathrm{S}_0}$,\;
$0 \leqslant {y_j}_{k_0k_1k_2}<{\mathrm{S}_1}$, \;
$0 \leqslant {z_j}_{k_0k_1k_2}<{\mathrm{S}_2}$ \; \\
\phantom{a way that \; }%
$(k_l=1,2,\ldots,m_l,\; l=0,1,2,\;\mathrm{S}_0=\mathrm{S}_1=\mathrm{S}_2=1)$,\\
\phantom{way that \; }%
$0 \leqslant {x_j}_{k_0k_1k_2}<{\mathrm{L}_0}$, \;
$0 \leqslant {y_j}_{k_0k_1k_2}<{\mathrm{L}_1}$, \;
$0 \leqslant {z_j}_{k_0k_1k_2}<{\mathrm{L}_2}$ \; \\
\phantom{way that \; }%
$(k_l=m_l+1,\ldots,K_l,\; l=0,1,2,\; \mathrm{L}_l=1+1/\chi_l)$.\\
As an example, consider 2D and 3D quasicrystals whose corresponding
probabilities $p_1(n+\beta),\, p_2(n+\beta)$ are defined
by formula (\ref{math/5}).
These quasicrystals will consist of four structural cells in the case of a 2D
quasicrystal, and eight ones in the case of a 3D quasicrystal.
Make a more detailed analysis of 2D quasicrystals. Introduce the probabilities
\begin{subequations}
\label{math/25}
\begin{align}
p_a(\mathbf{n}+\boldsymbol{\beta})=p_1(n_0+\beta_0)\,p_1(n_1+\beta_1)\,,\\
p_b(\mathbf{n}+\boldsymbol{\beta})=p_1(n_0+\beta_0)\,p_2(n_1+\beta_1)\,,\\
p_c(\mathbf{n}+\boldsymbol{\beta})=p_2(n_0+\beta_0)\,p_1(n_1+\beta_1)\,,\\
p_d(\mathbf{n}+\boldsymbol{\beta})=p_2(n_0+\beta_0)\,p_2(n_1+\beta_1)\,
\phantom{.}
\end{align}
\end{subequations}
and structural cells $A,\: B\: ,C\: ,D$ corresponding to these probabilities.
Consider the probabilities as functions of continuous variables $x,y$.
These functions are periodic in $x$ with the period $\sigma_0$,
 in $y$ with the period $\sigma_1$.
To demonstrate, plot the graphs of these functions in the plane and point out
the areas in which probabilities $p_a,p_b,p_c,p_d$ assume values equal to
unity (Fig.\ref{f:6}).
As seen from the graph:
\begin{subequations}
\label{math/26}
\allowdisplaybreaks
\begin{align}
p_a(x,y)=1,& \quad \mbox{if }\; \left|\begin{array}{l}
 n_0\sigma_0\leqslant{x}<\left(n_0 +1\right)\sigma_0 -1 ,\\
 n_1\sigma_1\leqslant{y}<\left(n_1 +1\right)\sigma_1 -1 ,
\end{array}\right. \vphantom{ \cfrac{1}{ 1+\frac{1}{ 1+\frac{1}{1} } } } \\
p_b(x,y)=1,& \quad \mbox{if }\; \left| \begin{array}{l}
 n_0\sigma_0\leqslant{x}<\left(n_0 +1\right)\sigma_0 -1 ,\\
(n_1+1)\sigma_1-1\leqslant{y}<\left(n_1 +1\right)\sigma_1 ,
\end{array}\right. \vphantom{ \cfrac{1}{ 1+\frac{1}{ 1+\frac{1}{1} } } } \\
p_c(x,y)=1,& \quad \mbox{if }\; \left| \begin{array}{l}
 \left(n_0+1\right)\sigma_0-1\leqslant{x}<\left(n_0 +1\right)\sigma_0,\\
 n_1\sigma_1\leqslant{y}<\left(n_1 +1\right)\sigma_1-1 ,
\end{array}\right.\vphantom{ \cfrac{1}{ 1+\frac{1}{ 1+\frac{1}{1} } } } \\
p_d(x,y)=1,& \quad \mbox{if }\; \left| \begin{array}{l}
 \left(n_0+1\right)\sigma_0-1\leqslant{x}<\left(n_0 +1\right)\sigma_0 ,\\
 \left(n_1+1\right)\sigma_1-1\leqslant{y}<\left(n_1 +1\right)\sigma_1 .
\end{array}\right.
\end{align}
\end{subequations}
Taking account of the periodicity $p_a$,~$p_b$,~$p_c$,~$p_d$ instead of the
conditions (\ref{math/26}) on the variables $x$,~$y$ throughout the plane $XY$,
one can consider conditions on the values
\begin{equation}
\label{math/27}
\begin{split}
&\Delta_0(x)=x-\sigma_0 \left\lfloor \frac{x}{\sigma_0} \right\rfloor=
\sigma_0\left\{\frac{x}{\sigma_0}\right\}
\vphantom{ \cfrac{1}{ 1+\frac{1}{ 1+\frac{1}{1}}}} \,,\\
&\Delta_1(y)=\sigma_1\left\{\frac{y}{\sigma_1}\right\}\\
\end{split}
\end{equation}
\pagebreak[1] in the rectangle $\sigma_0 \times \sigma_1$ placed in initial of coordinate.\\
Then formula (\ref{math/26}) can be written in the form:
\begin{subequations}
\label{math/28}
\begin{align}
p_a(x,y)=1,& \quad \mbox{if }\; 0\leqslant\Delta_0(x)<\sigma_0 -1,\:
 0\leqslant\Delta_1(y)<\sigma_1 -1 \,,\\
p_b(x,y)=1,& \quad \mbox{if }\; 0\leqslant\Delta_0(x)<\sigma_0 -1,\:
 \sigma_1-1\leqslant\Delta_1(y)<\sigma_1 \,,\\
p_c(x,y)=1,& \quad \mbox{if }\; \sigma_0 -1\leqslant\Delta_0(x)<\sigma_0,\:
 0\leqslant\Delta_1(y)<\sigma_1 -1\,,\\
p_d(x,y)=1,& \quad \mbox{if }\; \sigma_0 -1\leqslant\Delta_0(x)<\sigma_0,\:
 \sigma_1-1\leqslant\Delta_1(y)<\sigma_1 \,.
\end{align}
\end{subequations}
The probabilites $p_a(\mathbf{n}+\boldsymbol{\beta})$,
$p_b(\mathbf{n}+\boldsymbol{\beta})$, $p_c(\mathbf{n}+\boldsymbol{\beta})$,
$p_d(\mathbf{n}+\boldsymbol{\beta})$ and conditions for them,
analogous to (\ref{math/27}), we will obtain at $x=n_0+\beta_0$,~$y=n_1+\beta_1$.
The quasicrystal, thus assigned for a particular case of
$\sigma_0=\sigma_1=\tau$, is shown in Fig.\ref{f:7}.
In increasing the number of structural cells, analogously to (\ref{math/16}),
each region in the graph Fig.\ref{f:6} will be divided into several
rectangulars, within which, the corresponding probabilities will assume the
value equal to unity (Fig.\ref{f:8}).
As well as in a one-dimensional case (formulae \ref{math/22}, \ref{math/23}),
reciprocal location of the bases can be correlated. Using the properties of
probabilities, described in Section \ref{c1dqc}, one can construct a
sufficiently broad class of the assigned quasiperiodic structures, although in
this case one may have to take account of an infinite number of products  of
probabilities, analogous to (\ref{math/22}), correlating reciprocal location of
the bases. In the above considered technique of describing quasicrystals for a
definite cell, a corresponding probability had the form of a product of
one-dimensional probabilities. It is not always that with such a presentation
of probabilities one can describe quasiperiodic structures. In a more general
case, probabilities will be corresponded by regions of an arbitrary form
restrained by curves (Fig.\ref{f:9}).
Analytical presentation of such quasiperiodic structures will be the following:
\begin{subequations}
\label{math/29}
\allowdisplaybreaks
\begin{align}
p_k(\mathbf{n}+\boldsymbol{\beta})&=\left| \begin{array}{ll}
1,& \; \mbox{if }\: \left(\Delta_0\left(n_0+\beta_0\right),
\Delta_1\left(n_1+\beta_1\right)\right)\in{U_k} , \\
0,& \; \mbox{if }\: \left(\Delta_0\left(n_0+\beta_0\right),
\Delta_1\left(n_1+\beta_1\right)\right)\notin{U_k} ,
\end{array} \right.\\
\rho(\mathbf{n},\boldsymbol{\beta})&
=\sum_{n_0,\,n_1 = -\infty}^{\infty} \sum_{k=1}^{K}
p_k(\mathbf{n}+\boldsymbol{\beta})\sum_{j_k=1}^{J_k}
\delta(\mathbf{r-r_n(\boldsymbol{\alpha},{\boldsymbol\beta})-r}_{j_k}).
\end{align}
\end{subequations}
Such a technique of describing allows one by a quasiperiodic ornament  with the
assigned matching rule to restore all the probabilities of
$p_k(\mathbf{n}+\boldsymbol{\beta})$. To understand better, how by the assigned
matching to construct a quasicrystal, consider a particular example.
\subsection   {Example: the Penrose Matching}
\begin{figure}[tb]
\centering
\includegraphics*{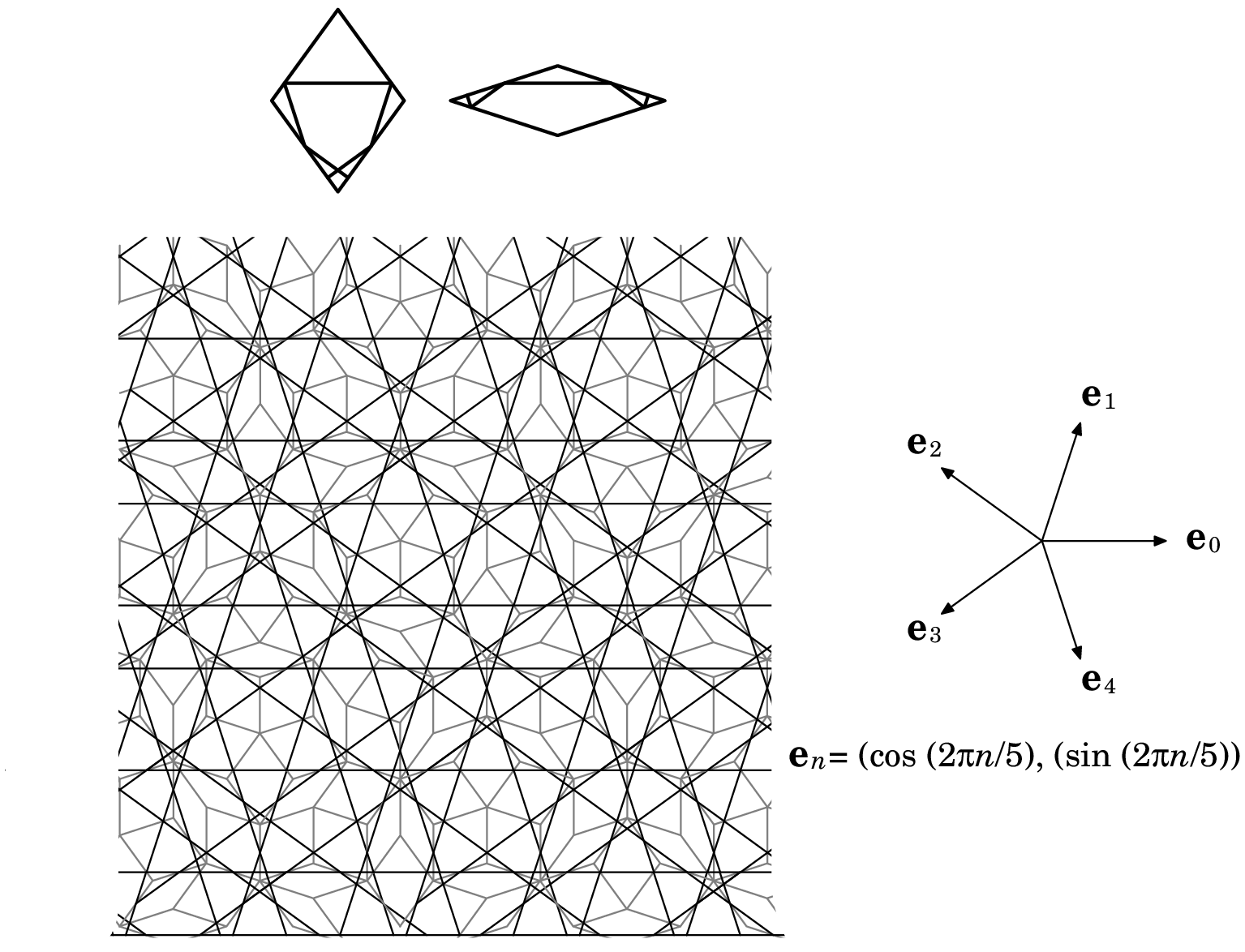}
\captn{Ammann line decoration of a portion of a Penrose tiling. The tiles
have been decorated as shown at top to reveal the Ammann quasilattice
associated with the tiling.}
\label{f:10}
\end{figure}

\begin{figure}[tb]
\includegraphics*{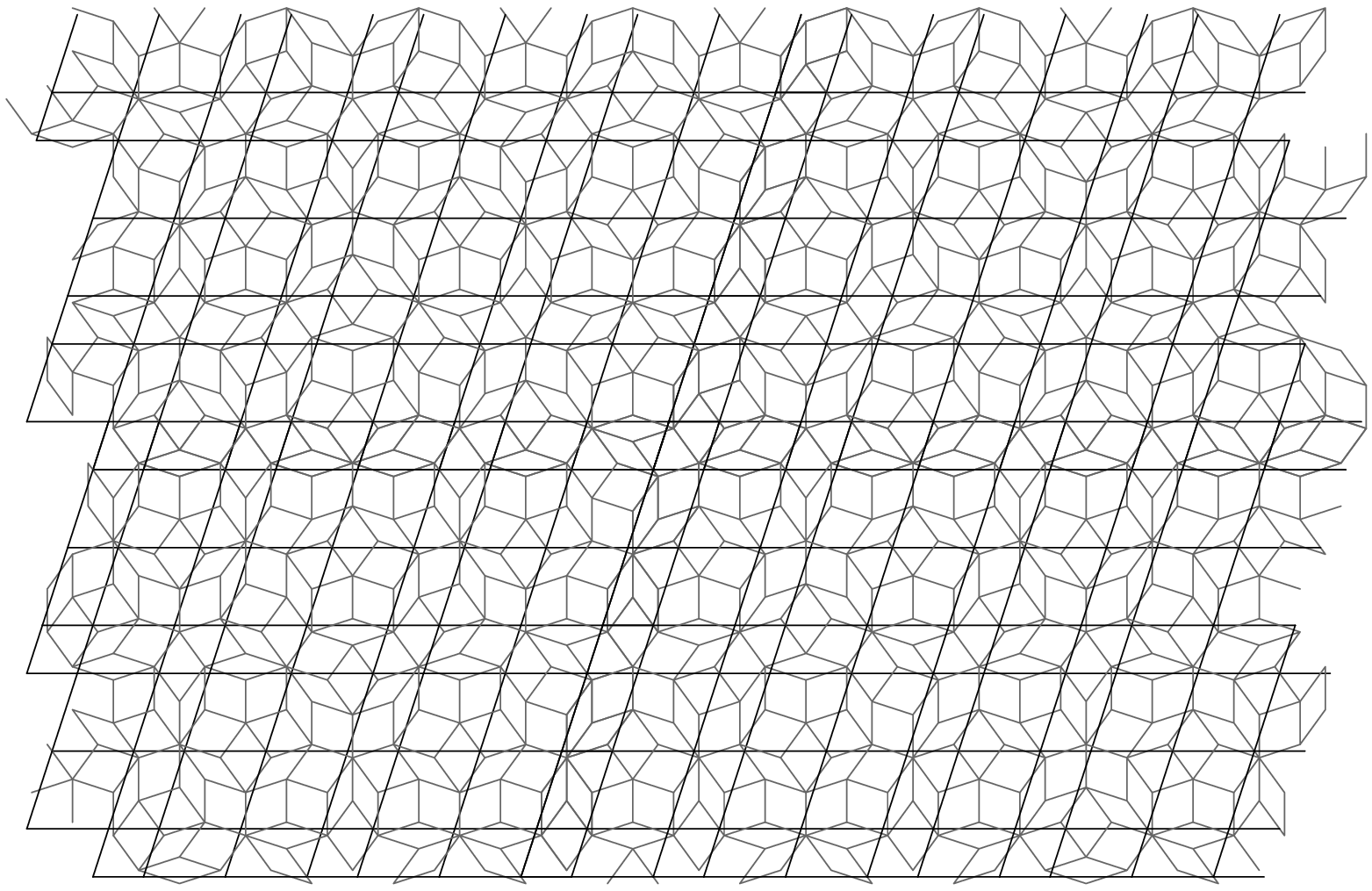}
\captn{Ammann lattice, intersecting the Penrose tiling, with reserved
lines parallel to $\mathbf{e}_0$,~$\mathbf{e}_1$. In such a quasiperiodic
lattice cell can be seen as a mosaic element covered with a certain pattern.}
\label{f:11}
\end{figure}

\begin{figure}[ptb]
\includegraphics*{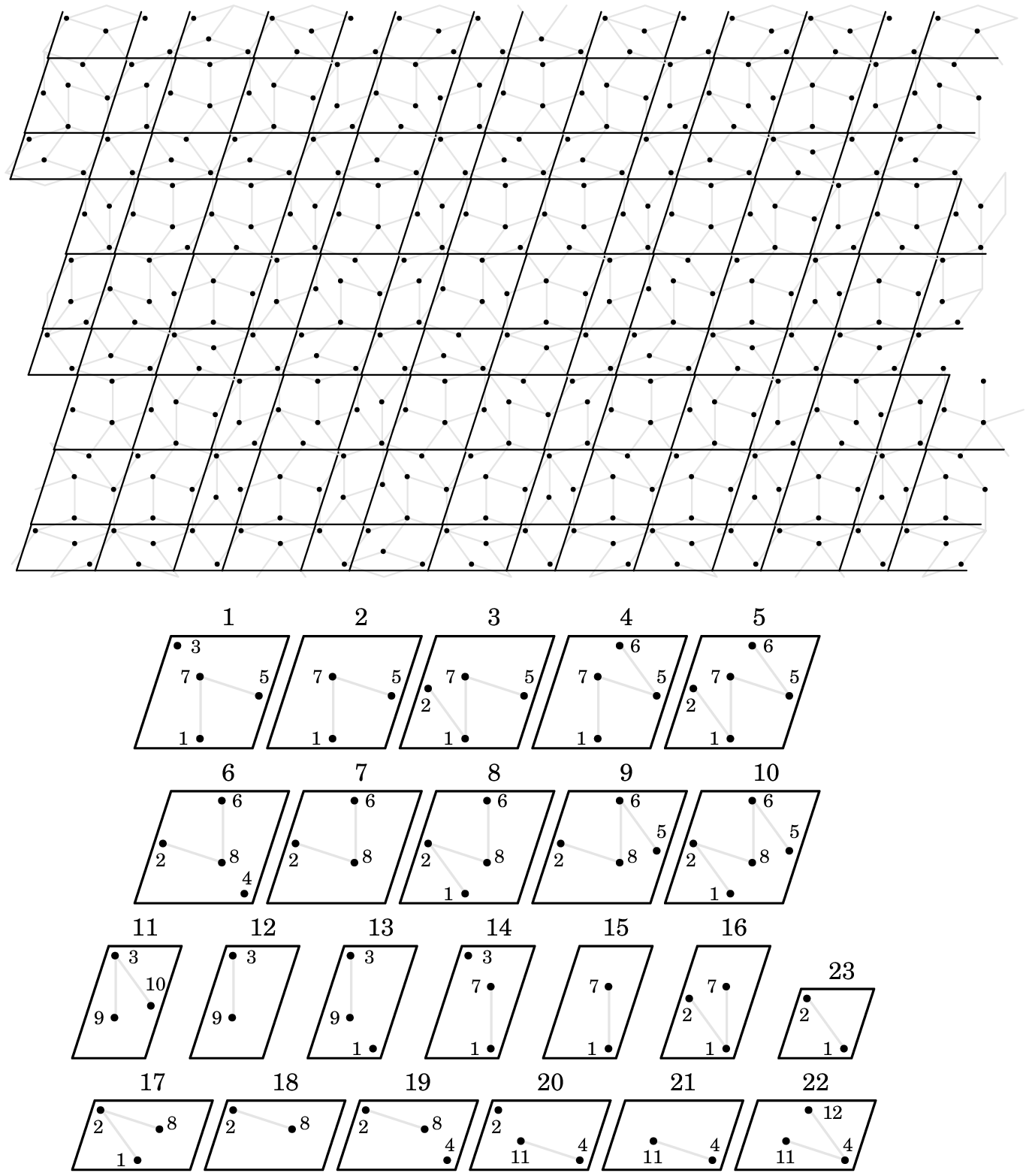}
\captn{23 bases of a Penrose quasicrystal consist of as few as 12 atoms.}
\label{f:12}
\end{figure}

\begin{figure}[ptb]
\centering

\includegraphics*{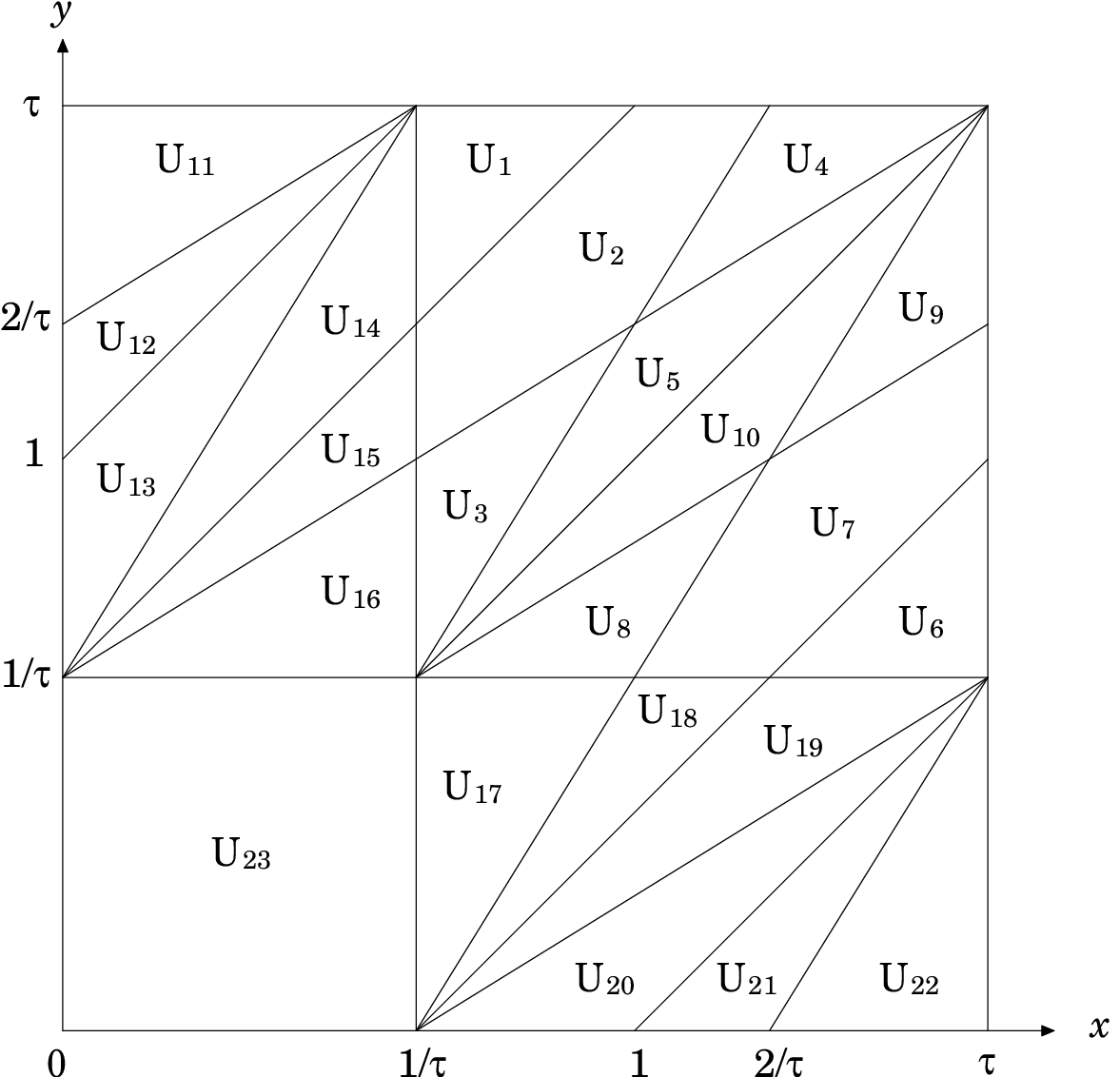}

\captn{The probability distribution graph of the Penrose quasicrystal,
in which the areas U$_k$, corresponding to definite bases.}
\label{f:13}
\end{figure}

To construct a quasiperiodic ornament consisting of two rhombi, whose sharp
angles are equal to $36^\circ$ and $72^\circ$, the rhombi are decorated with
sections of straight lines [9] (Fig .\ref{f:10}) and a rule is assigned  that in
making mosaic, sections on the rhombi are to form continuous lines. With such a
matching rule a quasiperiodic Penrose ornament is produced, which intersects a
pentagrid~[9] (Fig.\ref{f:10}) known as the Ammann quasilattice. Otherwise one
can say that a Penrose ornament, constructed from rhombi, decorates the Ammann
lattice or is a matching on the Ammann lattice.
The position of parallel lines in the Ammann lattice is given by the formula
[9]:
\begin{equation}
\label{math/30}
x_{n_l}=n_l+\alpha_l  +
 \frac{1}{\tau} \left\lfloor \frac{n_l+\beta_l}{\tau} \right\rfloor.
\end{equation}
Construct a Penrose matching on a beforehand assigned quasiperiodic lattice
(i.e. a with the assigned $\alpha_0$, $\alpha_1$, $\beta_0$, $\beta_1$),
consisting of a set of lines parallel to the vectors
$\mathbf{e}_0$, $\mathbf{e}_1$ so that the lines being formed from the sections,
decorating rhombi, should coincide with the assigned lines. In the Ammann
lattice, intersecting the Penrose tiling, eliminate the lines parallel to
$\mathbf{e}_2$,~$\mathbf{e}_3$,~$\mathbf{e}_4$. In Fig.\ref{f:11} such a
quasiperiodic lattice, decorated with a Penrose mosaic, is shown.
Each lattice cell can be seen as a mosaic element covered with a certain
pattern. A quasiperiodic structure, corresponding to the Penrose matching,
will be produced, if in each site of the matching one atom is located
(Fig.\ref{f:12}).
Such a quasicrystal consists of 23 bases (Fig.\ref{f:12}). Note that 23 bases consist
of as few as 12 atoms. One can find probabilities of each basis in two ways:
The first way is to correlate reciprocal location of quasicrystals bases using
the properties of probabilities multiplication, considered in Section 1.
The second way follows from the description principle itself of a quasicrystal.
To define a probability of any basis, with the assigned $n_0$ and $n_1$, calculate
$\Delta(n_0+\beta_0)$ and $\Delta(n_1+\beta_1)$, then the set of the points
($\Delta(n_0+\beta_0)$, $\Delta(n_1+\beta_1)$) for each basis will
thickly fill a certain region in the probability distribution graph.
In calculating by the  both ways the probability distribution graph will be of
the form shown in Fig.\ref{f:13}.

\begin{figure}[ptb]
\centering

\includegraphics{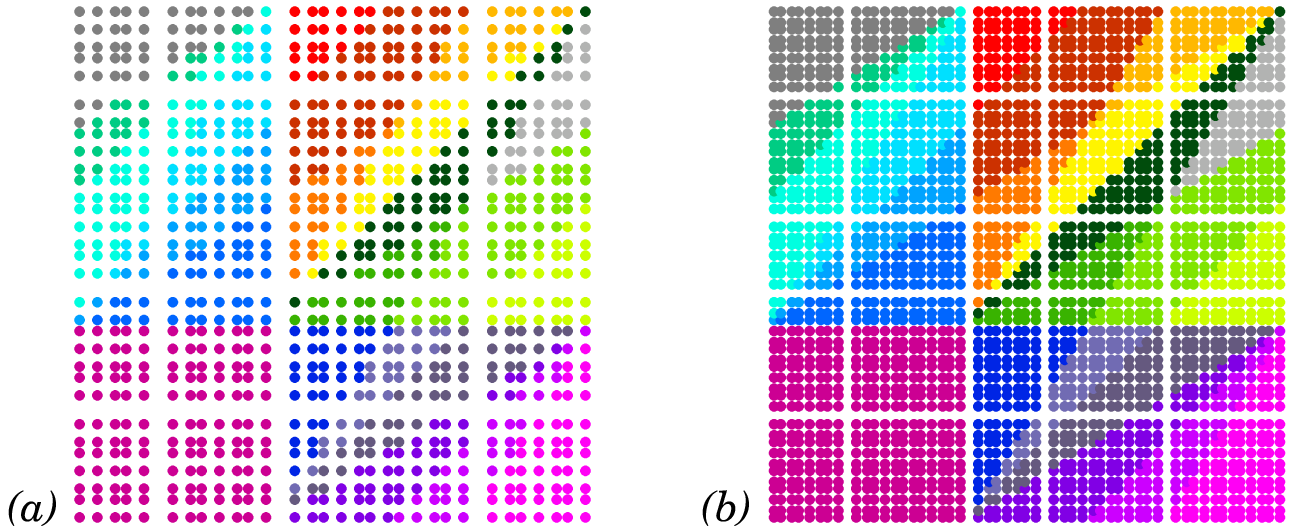}

\captn{\! The probability distribution graphs have been obtained with a numerical
analysis of the Penrose quasicrystal. $(a)$~30~by~30~cells, $(b)$~50~by~50~cells.
Points of one colour correspond to a definite basis.}
\label{f:14}
\end{figure}
\begin{figure}[htb]
\centering

\includegraphics{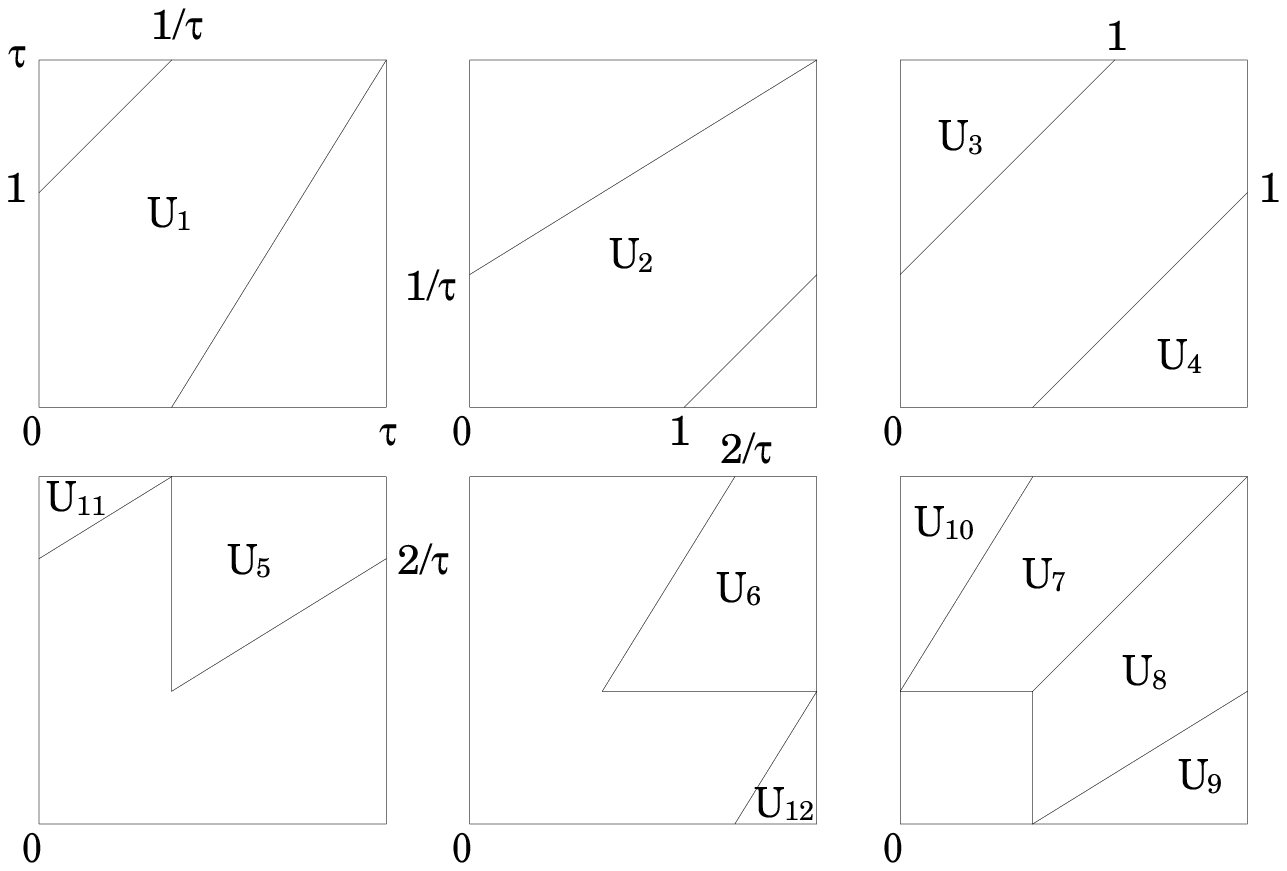}

\captn{The probability distribution graph of the Penrose quasicrystal, in which
the areas U$_k$, corresponding to definite atoms.}
\label{f:15}
\end{figure}
\begin{figure}[htb]
\centering
\includegraphics{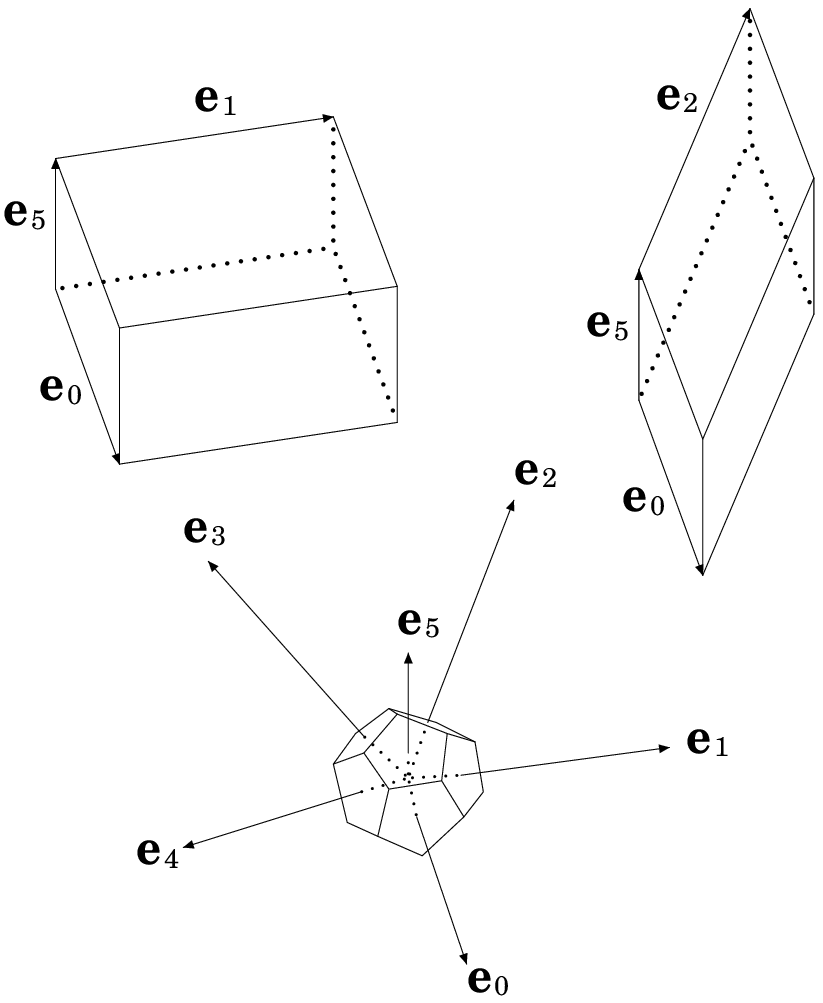}
\captn{Two types of rhombohedra that arise in icosahedral quasicrystal
packings.}
\label{f:16}
\end{figure}

Analytic presentation of the probabilities is of the form\footnote{Complete
analytic presentation of the probabilities is given in the Appendix}
(\ref{math/29}a).
Thus, the atoms density distribution $\rho(\mathbf{n},\boldsymbol{\beta})$ for
the Penrose quasicrystal can be written in the form of the sum
\begin{equation}
\label{math/31}
\rho(\mathbf{n},\boldsymbol{\beta})=\sum_{n_0{,}\:n_1=-\infty \:}^{\infty} \sum_{k=1}^{23}
p_k(\mathbf{n}+\boldsymbol{\beta})\sum_{j_k=1}^{J_k}
\delta(\mathbf{r-r_n(\boldsymbol{\alpha},{\boldsymbol\beta})-r}_{j_k}).
\end{equation}
Since in the Penrose crystal the locations of some atoms in different bases
coincide, one can find probabilities corresponding to each atom (altogether~12).
In Fig.\ref{f:15}  the regions U$_k$ are shown, in  which the probabilties,
corresponding to definite atoms, are equal to unity.
It is clear that, if several atoms belong to one basis, then the U$_k$ regions
of such atoms can overlap. In this interpretation analytic presentation of the
atoms density distribution $\rho(\mathbf{n},\boldsymbol{\beta})$ can be written
in the form of the sum :
\begin{equation}
\label{math/32}
\rho(\mathbf{n},\boldsymbol{\alpha},\boldsymbol{\beta})=
\sum_{n_0,\: n_1=-\infty \:}^{\infty} \sum_{j=1}^{12}p_j(\mathbf{n}+
\boldsymbol{\beta})\,\delta(\mathbf{r-r_n}(\boldsymbol{\alpha},
\boldsymbol{\beta})-\mathbf{r}_j)\,.
\end{equation}
Thus, on the axample of the Penrose mosaic the rules of constructing a
quasicrystal corresponding to the given matching have been obtained.
One can construct a quasicrystal by the assigned quasiperiodic matching only if
the following conditions are available: \\
(a) If there are several mosaic elements decorated with sections of straight
lines, so that in constructing a quasiperiodic matching from mosaic elements,
the  sections of straight lines form a grid consisting of $L$ groups of parallel
continuous lines.\\
(b) Position of lines  in a general case is assigned by the formula:
\begin{equation}
\label{math/33}
x_{n_l}(\alpha_l,\beta_l)=n_l+\alpha_l+\frac{1}{\chi_l}
\left\lfloor\frac{n_l+\beta_l}{\sigma_l} \right\rfloor
\end{equation}
where $1\leqslant l \leqslant L$, or
\begin{equation}
\label{math/34}
x_{n_l} = n_l + \alpha_l
\end{equation}
as this takes place, in the case of (\ref{math/34}) the $\sigma_l$ parameters
must be known. Then the order of constructing a quasicrystal by the assigned
matching will be the following:\\
(1) Reserve two groups of parallel lines from (\ref{math/33}) or (\ref{math/34}), and assign the
    parameters $\alpha_0, \alpha_1, \beta_0, \beta_1$ and also the main
    vectors of the translations $\mathbf{a}_0, \mathbf{a}_1$.\\
(2) On the assigned straight lines construct a quasiperiodic mathcing, so that
    the lines being formed from sections, decorating mosaic elements, should
    exactly coincide with the assigned straight lines.\\
(3) In the produced picture, eliminate the lines  parallel to
    $\mathbf{a}_2,\:\mathbf{a}_3,\:...\:\mathbf{a}_L$.\\
(4) Correlate a quasicrystal atom with every site of the produced matching.
    Then we get a set of several bases.\\
(5) In the produced (only graphically so far) quasicrystal, calculate a set of
    $\Delta_l(n_l+\beta_l)$ $(l=0,1)$ values for each basis. Subject to the
    condition that $-N_l \leqslant n_l \leqslant N_l$, and \mbox{$N_l \gg 1$},
    a  set of points with the coordinates
    $(\Delta_0(n_0+\beta_0),\, \Delta_1(n_1+\beta_1))$ for each basis will
    thickly fill a certain region in the probability distribution graph. \\
To demonstrate, we have considered a constructing of a quasicrystal on the
plane. Since computer modelling and analysis of three--dimensional lattices is
rather difficult, restrict ourselves only to the description of a rule for
constructing three dimensional quasicrystals. In the three--dimensional space,
three dimensional figures --- mosaic elements. For example, two rhombohedra
will be Penrose rombus analogues (Fig. \ref{f:16}) [9].
 \\
Three--dimensional mosaic elements are decorated with sections-areas of faces.
A generalization of a Penrose rule for a three-dimensional quasiperiodic
matching will be the following: in constructing a matching, the areas of the
planes, decorating three-dimensional mosaic elements, form continuous planes.
The planes are to constitute several groups of parallel intersecting planes.
Position of parallel planes is defined by the law (\ref{math/33}) or
(\ref{math/34}). In the three-dimensional case intersecting lines of the planes
will be crystallographic exes. For such a quasiperiodic matching, a
corresponding quasicrystal is constructed according to the following rule:\\
(1) Reserve three groups of parallel planes, whose position is determined by
    the law (\ref{math/33}) or (\ref{math/34}), and assign values of the
    parameters $\alpha_l$, $\beta_l$ and also the main vectors of the
    translations $\mathbf{a}_l \:(l=0,1,2)$.\\
(2) At given set of the planes, construct a quasiperiodic mathcing, so that
    the planes perpendicular to the vectors $\mathbf{a}_0 \times \mathbf{a}_1$,
    $\mathbf{a}_1 \times \mathbf{a}_2$, $\mathbf{a}_2 \times \mathbf{a}_0$
    formed from the plane areas, decorating three-dimensional mosaic elements,
    should exactly coincide with the assigned planes.\\
(3) In the produced volume picture eliminate the planes  non-perpendicular
    to the vectors  $\mathbf{a}_0 \times \mathbf{a}_1$,
    $\mathbf{a}_1 \times \mathbf{a}_2$, $\mathbf{a}_2 \times \mathbf{a}_0$ .\\
(4) Correlate a quasicrystal atom with every site of the produced matching. In
    this  case we obtain a set of several bases.\\
(5) In the produced (only graphically so far) quasicrystal, calculate a set of
    values $\Delta_l(n_l+\beta_l)$ $(l=0,\:1,\:2)$ . Subject to the  condition
    $-N_l \leqslant n_l \leqslant N_l$, and $N_l \gg 1$, a set of points with
    the coordinates %
   $(\Delta_0(n_0+\beta_0),\: \Delta_1(n_1+\beta_1),\: \Delta_2(n_2+\beta_2))$
    for each basis will thickly fill a certain area of the probability
    distribution graph.\\
\section               {Difraction in quasicrystals}
Radiation intensity, scattered on a crystal, is described by the formula
\begin{equation}
\label{math/35}
 I( \mathbf{q},\: \boldsymbol{\alpha},\: \boldsymbol{\beta} )=
  \left| \frac{1}{V}  \int_V \,d\mathbf{r}\,\exp(-i\,\mathbf{q \! \cdot \! r})
  \rho(\mathbf{r}, \: \boldsymbol{\alpha},\: \boldsymbol{\beta}) \right|^2 =
 \bigl| \hat{\rho}(\mathbf{q},\: \boldsymbol{\alpha},\: \boldsymbol{\beta}) \bigr|^2,
\end{equation}
where V is the crystal volume,
$\hat{\rho}(\mathbf{q},\: \boldsymbol{\alpha},\: \boldsymbol{\beta})$ is the
fourier--image of density distribution of the atoms,
$\: \mathbf{q}=\mathbf{a}^{\ast}_0\, q_0+\mathbf{a}^{\ast}_1\,%
q_1+\mathbf{a}^{\ast}_2\, q_2$,
$\: \mathbf{a}^{\ast}_i=\mathbf{a}_j \times \mathbf{a}_k /%
[\mathbf{a}_i\!\cdot\!(\mathbf{a}_j \times \mathbf{a}_k)]$.\\
To calculate a fourie-image of the atoms density distribution for a quasicrystal,
present $x_{n_l}$ defined by formula (\ref{math/33}) as
\begin{equation}
\label{math/36}
x_{n_l} = n_l \left(1+\frac{1}{\chi_l \sigma_l} \right)- \frac{1}{\chi_l}
  \left\{ \frac{n_l+\beta_l}{\sigma_l} \right\} +
 \frac{\beta_l}{\chi_l \sigma_l}+\alpha_l.
\end{equation}
Then the fourie-image of the atoms density distribution will be of the
form
\begin{equation}
\label{math/37}
 \hat{\rho}(\mathbf{q},\: \boldsymbol{\alpha},\: \boldsymbol{\beta}))=
 \frac{1}{V} \sum_{n_0\!,\; n_1 \!,\; n_2\:=-\infty\;}^{\infty}
  p_k(\mathbf+\boldsymbol{\beta})\prod_{l=0}^{2} F(l,n_l)
  \sum_{j_k}^{J_k}\exp(i\,\mathbf{q \! \cdot \! r}_{j_k} ),
\end{equation}
where $F(l,\,n_l)=\exp(i\,q_ln_l\gamma_l+i\,q_l\eta_l - i\,q_l/\chi_l
      \{(n_l+\beta_l)/\sigma_l \} )$,\\ \phantom{**********}
$\,\gamma_l=1+1/(\sigma_l\chi_l)$, $\;\eta_l = \beta_l/(\sigma_l\chi_l)+\alpha_l$.
\\ Consider in the right-hand side of (\ref{math/37}) the following expression
\begin{equation}
\label{math/38}
 \Psi(\mathbf{n + \boldsymbol{\beta},\, q})=\prod_{l=0}^{2}
 \exp\left(-i\, \frac{q_l}{\chi_l}
 \left\{\frac{n_l+\beta_l}{\sigma_l} \right\}\right)
 \sum_{j_k=1}^{J_k}\exp \left(i\,\mathbf{q \!\cdot\! r}_{j_k}\right)\,.
\end{equation}
Since the $\Psi(\mathbf{x,q})$ function is periodic in terms of $\mathbf{x}$
with the period $(\sigma_0,\sigma_1,\sigma_2)$, then $\Psi(\mathbf{x,q})$ can
be expanded into a Fourie series in terms of $\mathbf{x}$ in the parallelepiped with the
sides $\sigma_0$, $\sigma_1$, $\sigma_2$.
\begin{equation}
\label{math/39}
 \Psi(\mathbf{x,\,q})= \sum_{m_0\!,\: m_1 \!,\: m_2\,=-\infty\: }^{\infty}
 c(\mathbf{m,q}) \exp \left(-i\,2\pi \sum_{l=0}^{2}
 \frac{m_lx_l}{\sigma_l} \right).
\end{equation}
The Fourie coefficients $c(\mathbf{m,q})$ are calculated as follows:
\begin{equation}
\label{math/40}
c(\mathbf{m,q}) = \frac{1}{V_\sigma}
\int_{V_\sigma} \, d\mathbf{x}\, \Psi(\mathbf{x,\,q})
 \exp \left(i\,2\pi\sum_{l=0}^{2}\frac{m_lx_l}{\sigma_l} \right),
\end{equation}
where $V_{\sigma}$ is the parallelepiped volume with the sides
$\sigma_0$, $\sigma_1$, $\sigma_2$.
Calculation of the integral in the right-hand side of (\ref{math/40}) is
reduced to the calculation of $K$ integrals from the regions $V_k$ of the
parallelepiped $V_{\sigma}$, in which the corresponding probabilities
$p_k(\mathbf x)$ take the values equal to unity. Then formula (\ref{math/40})
will assume the following form
\begin{equation}
\label{math/41}
c(\mathbf{m,q}) =
\sum_{j_k=1}^{J_k} c_k(\mathbf{m,q}) \exp \left(i\,\mathbf{q\! \cdot\!
 r}_{j_k}\right),
\end{equation}
where
\begin{equation}
\label{math/42}
c_k(\mathbf{m,q}) = \frac{1}{V_\sigma} \int_{V_k} d\mathbf{x}\,
\exp \left( i\,x_l
\sum_{l=0}^{2} \left( \frac{2\pi m_l}{\sigma_l}-\frac{q_l}{\sigma_l \chi_l} \right)
 \right).
\end{equation}
The fourie-image of  the atoms distribution function will be the following
\begin{equation}
\label{math/43}
\begin{split}
\tilde{\rho}(\mathbf q,\boldsymbol\alpha,\boldsymbol\beta)=
& \frac{1}{V} \sum_{\mathbf{n,\, m}}
\exp \left( i\, \sum_{l=0}^{2}\left( q_l(n_l\gamma_l+\eta_l )-
\frac{2\pi m_l}{\sigma_l}(n_l+\beta_l)  \right) \right)\\
 &\times\sum_{k=1}^{K}\sum_{j_k=1}^{J_k} c_k(\mathbf{m,q})
\exp(i\,\mathbf{q\!\cdot\! r}_{j_k})\,.
\end{split}
\end{equation}
Summing in $\mathbf n$ obtain
\begin{equation}
\label{math/44}
\begin{split}
\tilde{\rho}(\mathbf q,\boldsymbol\alpha,\boldsymbol\beta)=&\frac{1}{V}
\sum_{\mathbf{n,\, m}} \delta(\mathbf{q-q_{nm}})
\prod_{l=0}^{2} \frac{1}{\gamma_l} \exp \left( i\,q_l\eta_l-
i\,2\pi\frac{\beta_l m_l}{\sigma_l} \right)\\
 &\times\sum_{k=1}^{K}\sum_{j_k=1}^{J_k} c_k(\mathbf{m,q})
\exp(i\,\mathbf{q\!\cdot\! r}_{j_k}),
\end{split}
\end{equation}
where
\[ \mathbf{q_{nm}}=\sum_{l=0}^{2}\left( \frac{2\pi n_l}{1+1/\sigma_l \chi_l}+
\frac{2\pi m_l}{1+1/\chi_l}  \right)\mathbf{a}^{\ast}_l. \]
Using  formula (\ref{math/44}) write the Fourie series of
$\rho(\mathbf r,\boldsymbol{\alpha},\boldsymbol{\beta})$
\begin{equation}
\label{math/45}
\begin{split}
\rho(\mathbf r,\boldsymbol{\alpha},\boldsymbol{\beta})=&\sum_{\mathbf{n,\, m}}
\prod_{l=0}^{2}
\exp \left( i\,q_l\eta_l-i\,2\pi\frac{\beta_l m_l}{\sigma_l} \right)\\
 &\times\sum_{k=1}^{K}\sum_{j_k=1}^{J_k} c_k(\mathbf{m,q})
\exp \left(i\,\mathbf{q\!\cdot\! (r+r}_{j_k})\right)\,.
\end{split}
\end{equation}
Then radiation intensity scattered on a quasicrystal can be written as
\begin{multline}
\label{math/46}
\allowdisplaybreaks
I(\mathbf{q},\boldsymbol{\alpha},\boldsymbol{\beta}) = \frac{1}{V} \int_V
\exp(i\,\mathbf{q\! \cdot \!r})
\rho(\mathbf{r},\boldsymbol{\alpha},\boldsymbol{\beta})
\hat{\rho}(\mathbf{q},\boldsymbol{\alpha},\boldsymbol{\beta})\\
\shoveleft{\quad =\frac{1}{V^2}\sum_{\begin{subarray}{c}
\mathbf{n}^{\phantom\prime},\, \mathbf{m}^{\phantom\prime}\\
\mathbf{n}^{\prime},\, \mathbf{m}^{\prime} \end{subarray}}
\sum_{k,\,{k^\prime}}^{K,\,{K^\prime}}
\sum_{j_{k^{\phantom{\prime}}} \! , \, j^{\,\prime}_{k^\prime}}
^{J_{k^{\phantom{\prime}}} \! , \, J_{k^\prime}}
c_k(\mathbf{m,q})c_{k^{\prime}}(\mathbf{m^{\prime},q})
\delta(\mathbf{q-q_{nm}})}\\
\times\exp \left( i\, \mathbf{q} \!\cdot\!
(\mathbf{r}_{j^{\phantom{a}}_k}-\mathbf{r}_{j^{\,\prime}_{k^\prime}})
+ i\, \sum_{l=0}^{2} 2\pi \frac{\beta_l}{\sigma_l}(m_{l^{\prime}}-m_l)\right)\\
\times\int_V d \mathbf{r} \exp \left( i\, \mathbf{r\!\cdot\! \left(
q_{n^{\prime}m^{\prime}}-q_{nm} \right)} \right).
\end{multline}
Since $V \gg 1$ the integral in the right-hand side of (\ref{math/46}) has the
following values
\begin{equation}
\label{math/47}
\int_V d \mathbf{r} \exp \left( i\, \mathbf{r\!\cdot\! \left(
q_{n^{\prime}m^{\prime}}-q_{nm} \right)} \right)=\left|
\begin{array}{ll}
V\, , &  \mbox{if } \quad \mathbf{q_{n^{\prime}m^{\prime}}-q_{nm}}=0\\
0\, , &  \mbox{if } \quad \mathbf{q_{n^{\prime}m^{\prime}}-q_{nm}}\not=0\, .
\end{array} \right.
\end{equation}
\begin{figure}[htb]
\centering
\includegraphics*{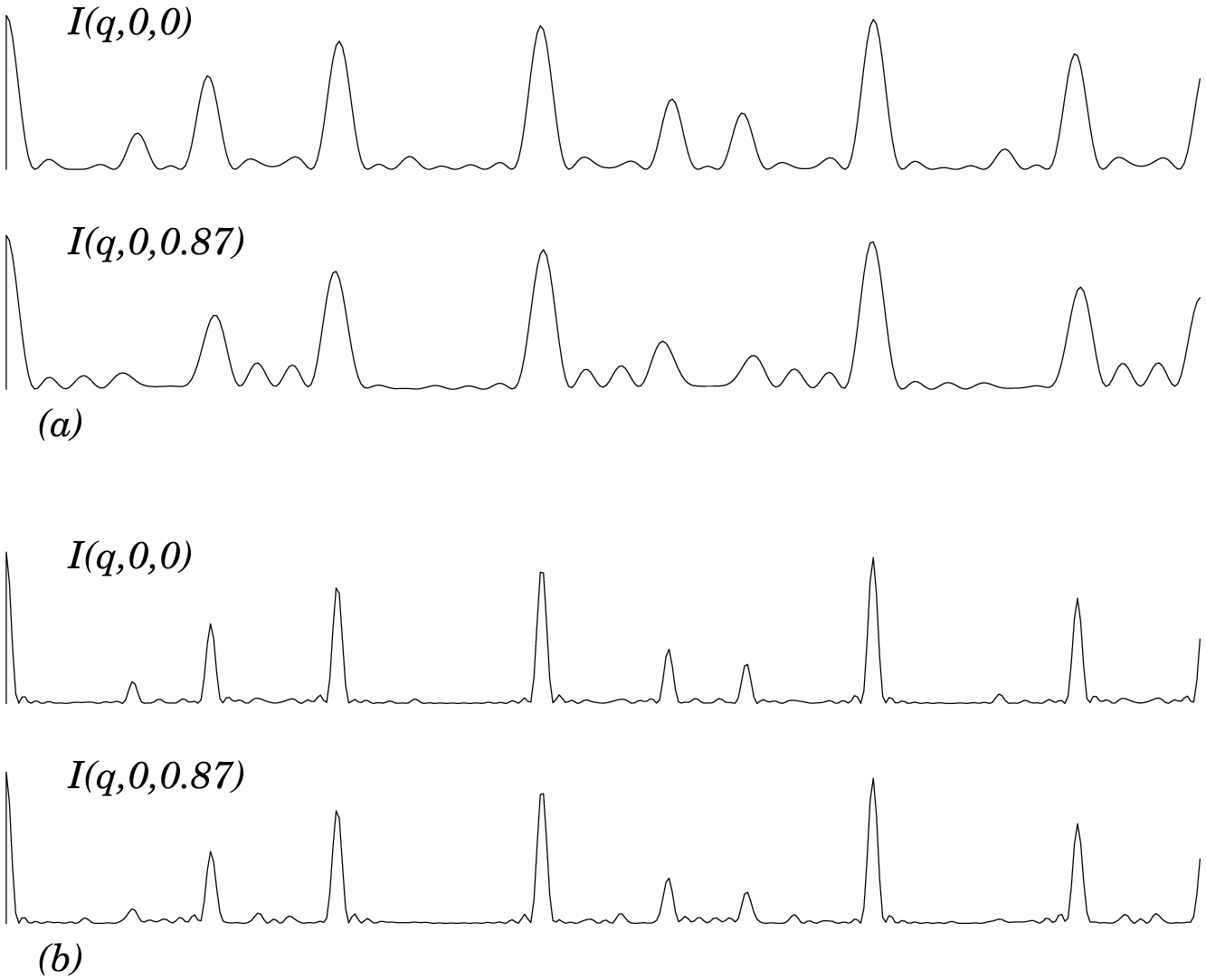}
\captn{\!Fourie-spectrum of a quasiperiodic sequence of $\delta$ functions, assigned
by formula (\ref{math/7}), $(a)$ $n=7$, $(b)$ $n=17$.}\label{f:17}
\end{figure}
I.e.
\begin{equation}
\label{math/48}
\int_V d \mathbf{r} \exp \left( i\, \mathbf{r\!\cdot\! \left(
q_{n^{\prime}m^{\prime}}-q_{nm} \right)} \right)=
V\prod_{l=0}^{2} \delta_{n^{\prime}_l n_l} \delta_{m^{\prime}_l m_l}.
\end{equation}
Thus, the intensity of the scattered radiation is not equal to zero, when
$\mathbf{n=n^{\prime} ,\; m=m^{\prime}}$.
\begin{multline}
\label{math/49}
I(\mathbf{q},\boldsymbol{\alpha},\boldsymbol{\beta}) = \\
\shoveleft{\, =\frac{1}{V}\sum_\mathbf{n,\,m}
\sum_{k,\,{k^\prime}}^{K,\,{K^\prime}}
\sum_{j^{\phantom{\prime}}_{k^{\phantom{\prime}}}\!,\,j^{\,\prime}_{k^\prime}}
^{J_{k^{\phantom{\prime}}}\!,\,J_{k^\prime}}
c_k(\mathbf{m,q})c_{k^{\prime}}(\mathbf{m,q})\delta(\mathbf{q-q_{nm}})}
\exp \left( i\, \mathbf{q} \!\cdot\!
(\mathbf{r^{\phantom{a}}_{j^{\phantom{a}}_k}}
-\mathbf{r_{j^{\,\prime}_{k^\prime}}})
\right).
\end{multline}
The following result is produced: intensity of a scattered radiation on an
infinite crystal is independent of the $\boldsymbol{\alpha}$ and
$\boldsymbol{\beta}$ parameters
\begin{equation}
\label{math/50}
I(\mathbf{q},\boldsymbol{\alpha},\boldsymbol{\beta}) =
I(\mathbf{q},\boldsymbol{0},\boldsymbol{0}) = I(\mathbf{q}).
\end{equation}
\begin{figure}[htb]
\centering
\includegraphics{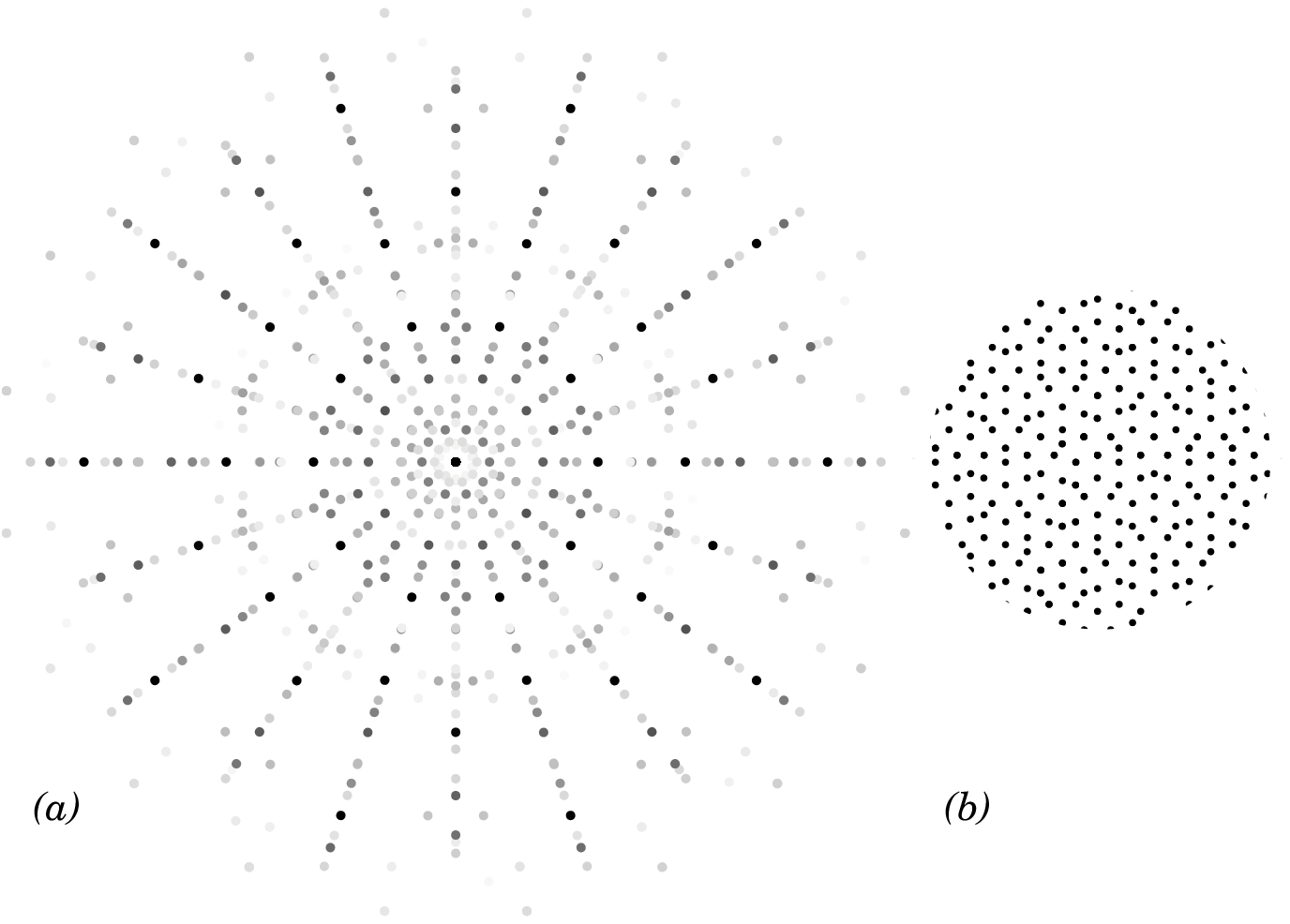}
\captn{The Fourie-spectrum $(a)$ of a Penrose quasicrystal $(b)$.}
\label{f:18}
\end{figure}
%
\section           {Symmetry Transformations for\\ Quasicrustals}
Since physical property of a quasicrystal, radiation scattering, is
independent of $\boldsymbol\alpha$ and $\boldsymbol\beta$, introduce an
invariance condition for a quasicrystal: if under the operator influence of
the transformation $\boldsymbol{\hat{\lambda}}$ on the atoms density
distribution $\rho(\mathbf{r},\boldsymbol{\alpha},\boldsymbol{\beta})$ only
the $\boldsymbol{\alpha}$ and $\boldsymbol{\beta}$ parameters change, but not
the $\rho$ function itself
\begin{equation}
\label{math/51}
\boldsymbol{\hat{\lambda}}
\rho(\mathbf{r},\boldsymbol{\alpha},\boldsymbol{\beta})=
\rho(\mathbf{r},\boldsymbol{\alpha}^{\,\prime},\boldsymbol{\beta}^{\,\prime})
\end{equation}
then the quasicrystal is invariant in respect to such a transformation.
\subsection         {Translational Invariance}
As has been shown above, a quasicrystal can be modelled with a quasiperiodic or
periodic lattice. Each site of the lattice is related
to some group of atoms---the basis, there being several such bases, and their
repetition frequency is determined by the $p_k(\mathbf{n}+\boldsymbol{\beta})$
probabilities. Let us see whether the quasicrystal, thus modelled,
is translationally invariant. Shift the lattice in such a way that the lattice
$\mathbf{m}$ site should be in the point with the coordinates $\boldsymbol{\alpha}$, then in the
lattice, produced as  result of the shift, the coordinate of any point will be
determined by the vector
\begin{equation}
\label{math/52}
\mathbf{r_n}(\boldsymbol{\alpha},\boldsymbol{\beta}^{\,\prime})=
\mathbf{r_{n+m}}(\boldsymbol{\alpha},\boldsymbol{\beta})
-\mathbf{r_{m}}(\boldsymbol{0},\boldsymbol{\beta}).
\end{equation}
For each component we obtain the following result
\begin{multline}
\label{math/53}
x_{n_l}(\alpha_l,\beta_l^{\,\prime})=\left( n_l +m_l+\alpha_l+
\frac{1}{\chi_l} \left\lfloor \frac{n_l+m_l+\beta_l}{\sigma_l} \right\rfloor \right)
-\left(m_l+ \frac{1}{\chi_l} \left\lfloor \frac{m_l+\beta_l}{\sigma_l}
\right\rfloor \right)
\vphantom{ \cfrac{1}{ 1+\frac{1}{ 1+\frac{1}{1}}}} \\
=n_l+\alpha_l+\frac{1}{\chi_l}
\left\lfloor \frac{n_l}{\sigma_l}+\frac{m_l+\beta_l}{\sigma_l}-
\left\lfloor \frac{m_l+\beta_l}{\sigma_l}\right\rfloor \right\rfloor=
\vphantom{ \cfrac{1}{ 1+\frac{1}{ 1+\frac{1}{1}}}} \\
n_l+\alpha_l+\frac{1}{\chi_l}
\left\lfloor \frac{n_l}{\sigma_l}+\left\{ \frac{m_l+\beta_l}{\sigma_l}\right\}
 \right\rfloor \,.
\end{multline}
I.e.
\begin{equation}
\label{math/54}
x_{n_l}(\alpha_l,\beta_l^{\,\prime})=
n_l+\alpha_l+\frac{1}{\chi_l}
\left\lfloor
 \frac{ n_l+\sigma_l \left\{ (m_l+\beta_l)/\sigma_l  \right\} } {\sigma_l}
\right\rfloor .
\end{equation}
The probabilities are the functions of the $\Delta_l(n_l+\beta_l)$ values for
which the equation
\begin{equation}
\label{math/55}
\Delta_l(n_l+m_l+\beta_l)
=\Delta_l(n_l+\sigma_l \left\{(m_l+\beta_l)/\sigma_l \right\})
\end{equation}
holds true.
Thus, in translating a quasicrystal get the following formula
\begin{equation}
\label{math/56}
T(\mathbf{m})\rho(\mathbf{r},\boldsymbol{\alpha},\boldsymbol{\beta})=
\rho(\mathbf{r},\boldsymbol{\alpha},\boldsymbol{\beta}^{\,\prime}),
\end{equation}
where $T(\mathbf{m})$ is the translation operator,
$\beta_l^{\,\prime}=\sigma_l\{(\beta_l+m_l)/\sigma_l \}$. Since the translation
transformation for a quasicrystal satisfies the condition (\ref{math/51}), a
quasicrystal is invariant in respect to translation transformations.
Consider an operation of translation multiplication
\begin{equation}
\label{math/57}
T(\mathbf{m}^{(2)})T(\mathbf{m}^{(1)})
\rho(\mathbf{r},\boldsymbol{\alpha},\boldsymbol{\beta})=
T(\mathbf{m}^{(2)})\rho(\mathbf{r},\boldsymbol{\alpha},\boldsymbol{\beta}^{\,(1)})
=\rho(\mathbf{r},\boldsymbol{\alpha},\boldsymbol{\beta}^{\,(1)(2)}).
\end{equation}
\begin{figure}[hbt]
\centering
\includegraphics*{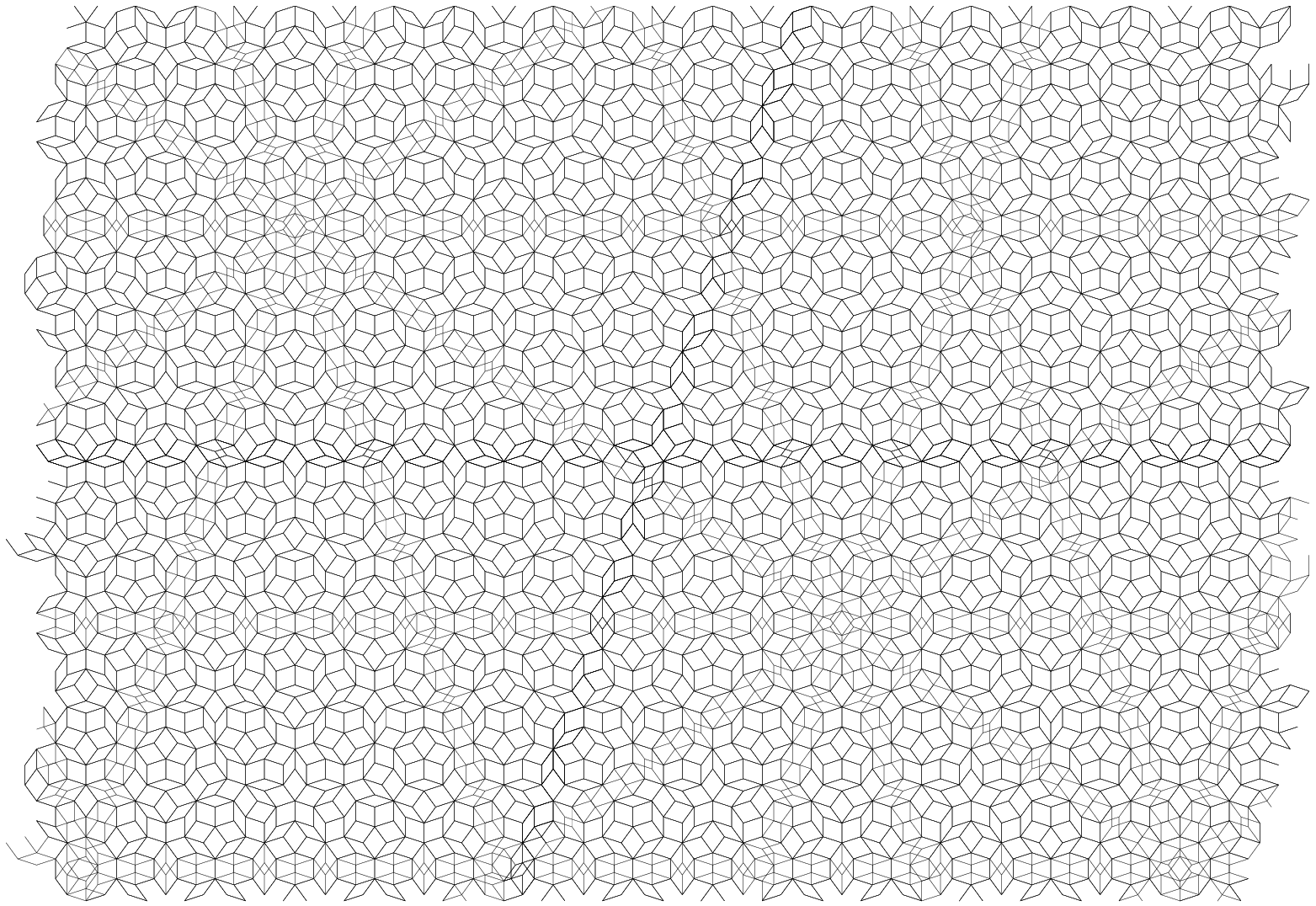}
\captn{Superposition of two Penrose mosaics: $\delta_1=0.12/\tau$,
$\delta_2=0.18/\tau$.} \label{f:19}
\end{figure}
Write out the $\boldsymbol{\beta}$ parameter transformation for an individual component
\begin{equation}
\label{math/58}
\begin{split}
\beta_l^{\,(1)}&=\sigma_l \left\{ \frac{\beta_l+m_l^{(1)}}{\sigma_l} \right\}
\vphantom{ \cfrac{1}{ 1+\frac{1}{ 1+\frac{1}{1+ frac{1}{1} }}}} \,,\\
\beta_l^{\,(1)(2)}&=\sigma_l \left\{ \frac{\beta_l^{\,(1)}+m_l^{(2)}}{\sigma_l} \right\}=
\sigma_l \left\{ \left\{ \frac{\beta_l+m_l^{(1)}}{\sigma_l} \right\}+
\frac{m_l^{(2)}}{\sigma_l} \right\}.
\end{split}
\end{equation}
For the $\{x\}$ function the equation $\{x\}=\{x+n\}$ $(n=\pm 1,\pm 2,\ldots)$
holds true, therefore the equation of the form $\bigl\{\{x\}+y\bigr\}$ can be
transformed as follows
\begin{equation}
\label{math/59}
\bigl\{\{x\}+y\bigr\}=\bigl\{\{x\}+\lfloor{x}\rfloor +y \bigr\}=
\bigl\{x+y\bigr\}
\end{equation}
it means that
\begin{equation}
\label{math/60}
\beta_l^{\,(1)(2)}=\sigma_l
\left\{ \frac{\beta_l+m_l^{(1)}+m_l^{(2)}}{\sigma_l} \right\}\,.
\end{equation}
Then for translation operators we get the following condition
\begin{equation}
\label{math/61}
T(\mathbf{m}^{(2)})\,T(\mathbf{m}^{(1)})=T(\mathbf{m}^{(2)}+\mathbf{m}^{(1)} )\,.
\end{equation}
Translation operators, corresponding to the property of (\ref{math/61})
form a cyclic group [17]. If in shifting the quasicrystal
$ 1 \gg \delta_l=|\beta_l/\sigma_l-{(\beta_l+m_l)/\sigma_l}|$,
large areas of the initial and the shifted quasicrystals will
coincide (Fig.\ref{f:19}).

\subsection          {Rotational Symmetries}
\begin{figure}[hbt]
\centering
\includegraphics*{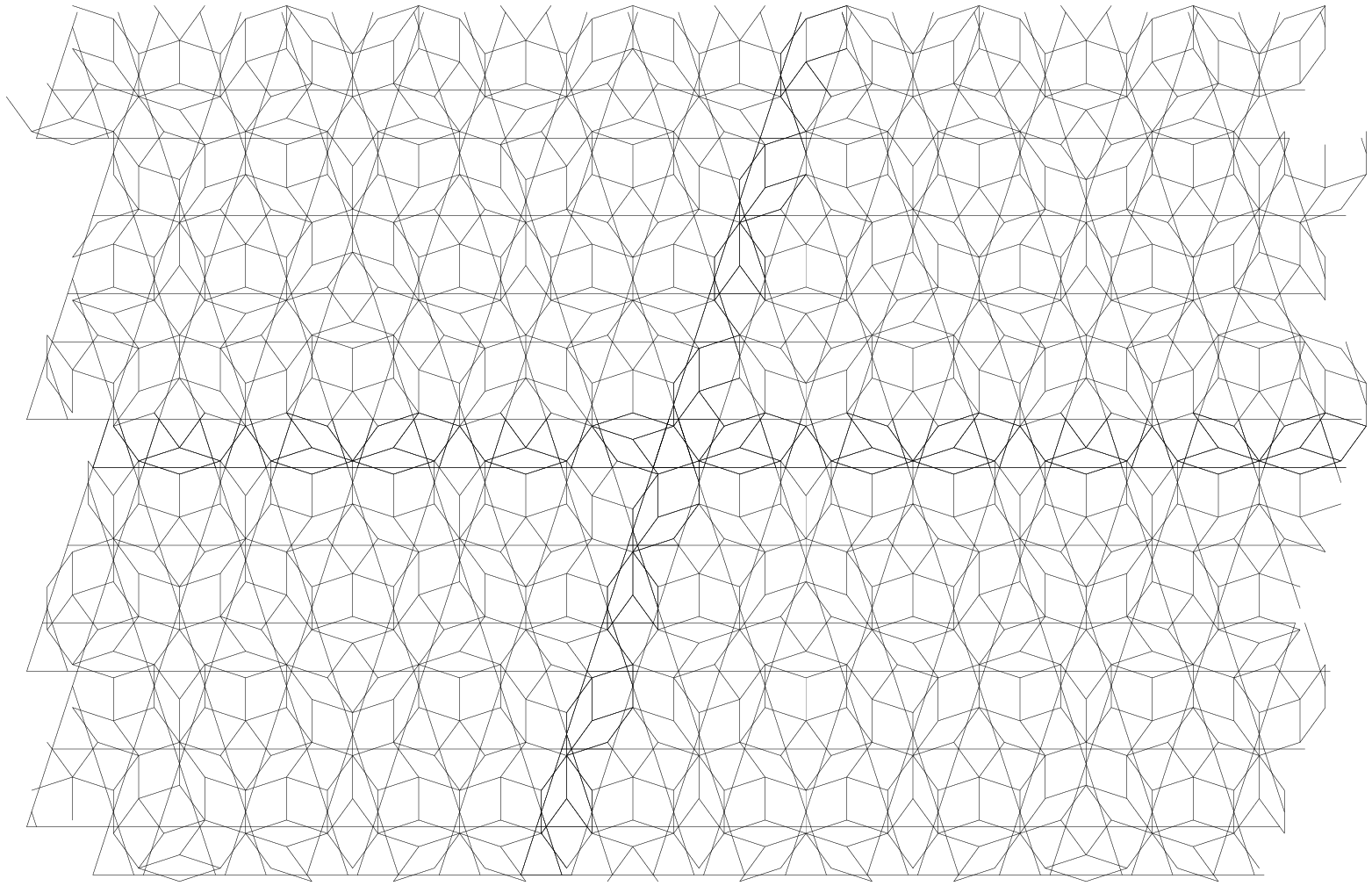}
\captn{The Ammann quasilattice, intersecting the Penrose tiling, with reserved
lines parallel to $\mathbf{e}_0$,~$\mathbf{e}_1$~$\mathbf{e}_4$ vectors.}
\label{f:20}
\end{figure}

\begin{figure}[hbt]
\centering
\includegraphics*{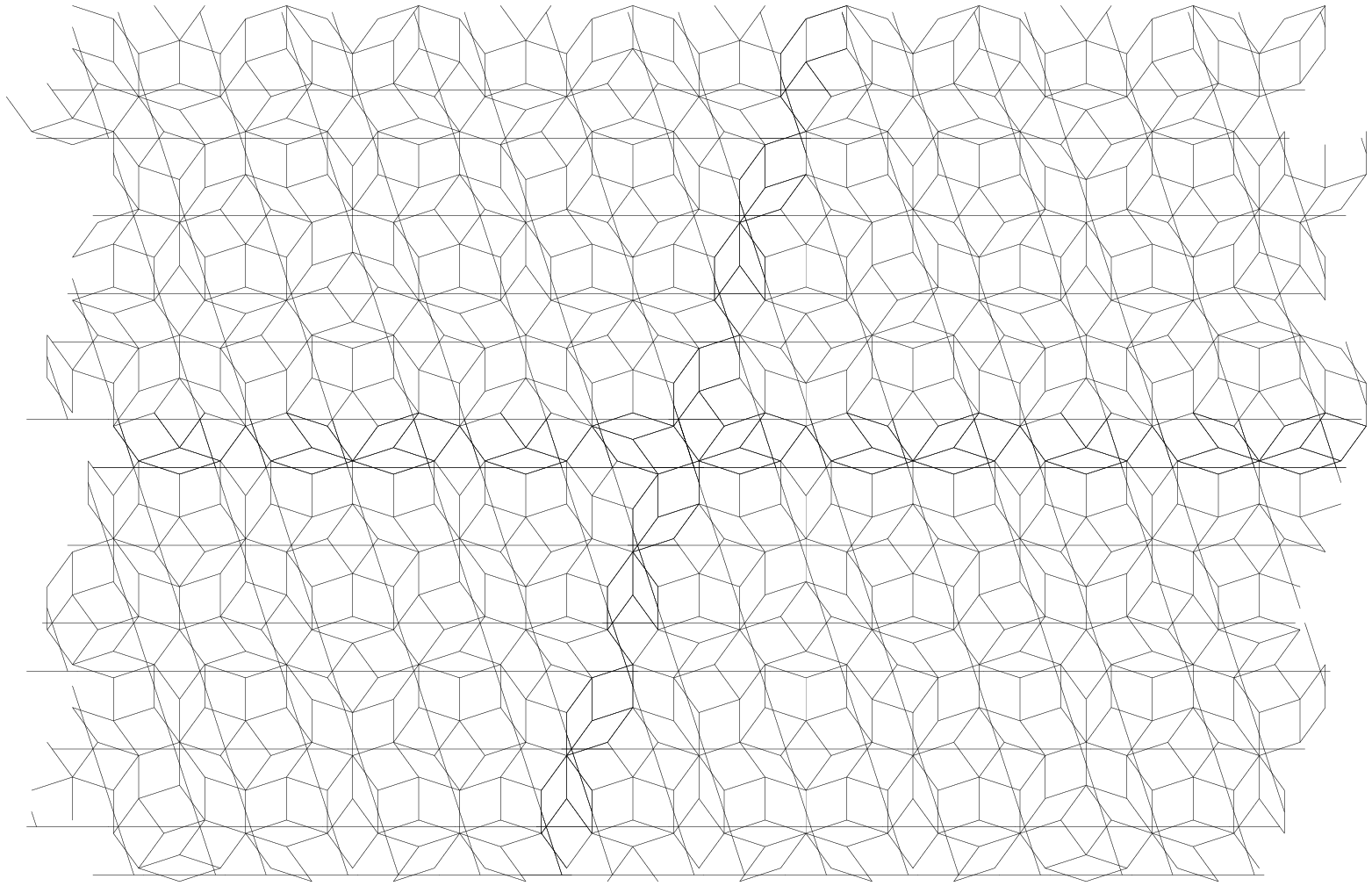}
\captn{The Ammann quasilattice, intersecting the Penrose tiling, with reserved
lines parallel to $\mathbf{e}_0$,~$\mathbf{e}_4$ vectors.}
\label{f:21}
\end{figure}

\begin{figure}[htb]
\centering
\includegraphics*{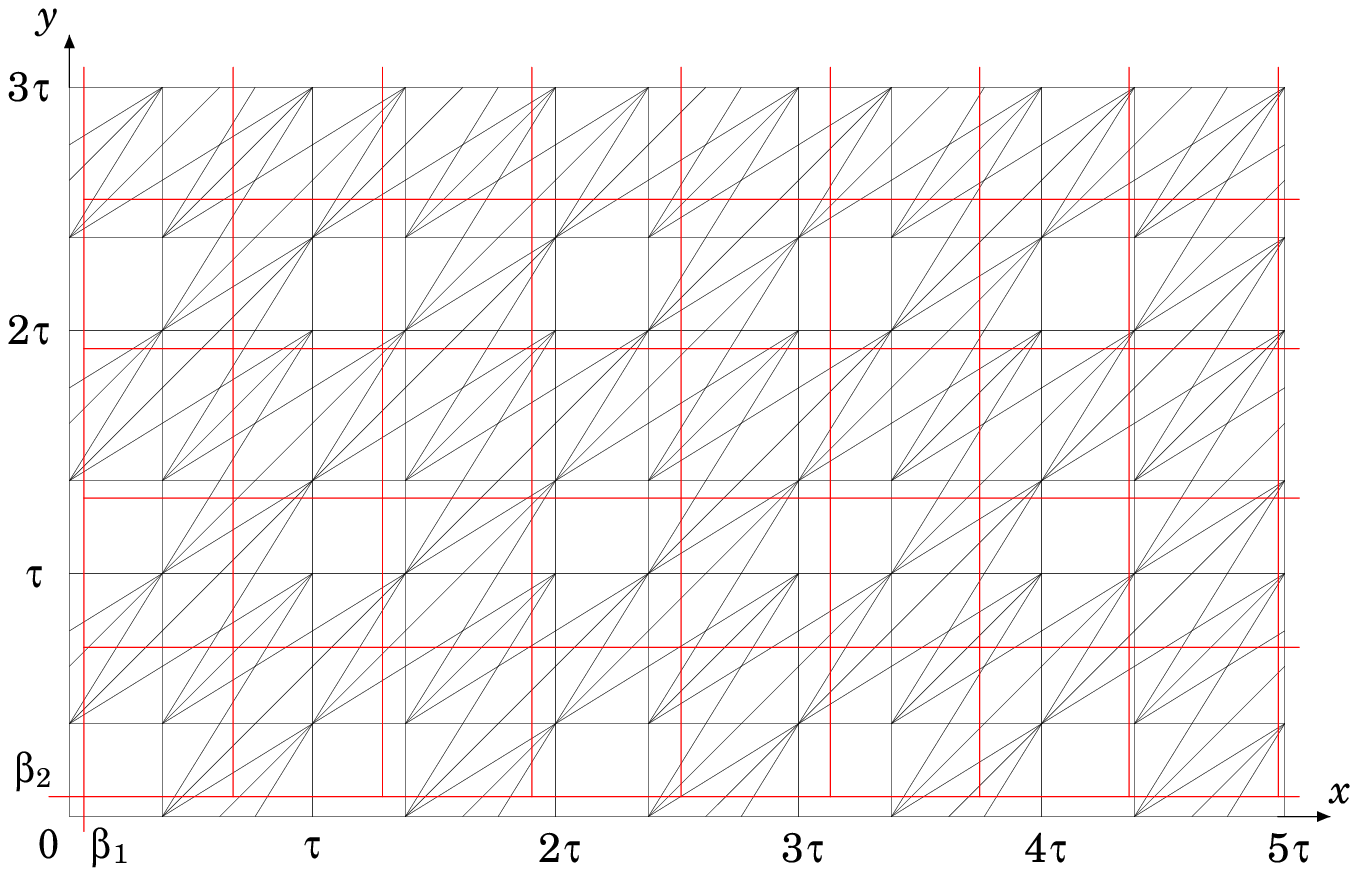}
\captn{A Penrose quasicrystal will be produced, if a two-dimensional square
lattice with a single step is superposed on the probability distribution graph,
periodically continued throughout the whole plane, than a point
of the lattice with the coordinates $(x, y)$ should be within the
U$_k$ region of a corresponding structural cell.}
\label{f:22}
\end{figure}

As has been mentioned earlier, quasicrystals can possess rotational symmetries
not allowed for ordinary crystals. Consider rotational symmetries on the
example of the Penrose matching. Reserve in the Ammann quasilattice,
intersecting the Penrose tiling, the lines parallel to the
$\mathbf{e}_0$, $\mathbf{e}_1$, $\mathbf{e}_4$ vectors (Fig.\ref{f:20}).
The Penrose quasicrystal can be considered in the coordinate axes $x_0$, $x_1$
(Fig.\ref{f:11}), or $x_0$, $x_4$ (Fig.\ref{f:21}).
In the both cases a crystal is produced  with the same structural cells and the
common law of their distribution, only the $\alpha_l$, $\beta_l$ parameters
will change. I.e. densities of the atoms distribution in the both cases have the
form (\ref{math/31}). Assign the values of parameters in the first case, since
any lattice site can be placed at the origin of the coordinates:
$\alpha_0=\alpha_1=0$,
$0 \leqslant \beta_0,\:\beta_1 < \tau$. To find parameters for a quasicrystal
considered in the exes parallel to $\mathbf{e}_0$, $\mathbf{e}_4$, select one
of the cells, partially superimposed with the crystal cell, constructed in the
axes, parallel to $\mathbf{e}_0$, $\mathbf{e}_1$ and being at the origin of
coordinates. After the cell has been selected, the parameters
$\alpha_0^{\,\prime}$, $\alpha_4^{\,\prime}$ are found geometrically. The
parameters $\beta_0^{\,\prime}$, $\beta_4^{\,\prime}$ are calculated by the
probability distribution graph (Fig.\ref{f:13}). The first cell corresponding
to the probability $p_k$, will give a value region of the parameters
$\beta_0^{\,\prime}$, $\beta_4^{\,\prime}$ within the region U$_{k^{\,\prime}}$.
On moving by a single
step in the direction of $\mathbf{e}_0$, or $\mathbf{e}_4$ from the region
U$_k$ one can get to several U$_{k^{\,\prime}}$ regions permissible for
realization of the Penrose quasicrystal. If on moving by a single step from the
U$_k$ region the moving is possible only to a definite U$_{k^{\,\prime}}$
region, corresponding to
a cell in the assigned quasicrystal, then the u$_{k^{\,\prime}}$ region of value
parameters will constrict, it is clear that u$_{k^{\,\prime}} \subset \mathrm{U}_{k^{\,\prime}}$.
Correlating a cell, selected at the origin of coordinates with severel cells of
the quasicrystal will reduce the u$_k$ region of value parameters
$\beta_0^{\,\prime}$, $\beta_4^{\,\prime}$. The described technique to determine
parameters can be presented as follows: consider a probability distribution graph
of a Penrose quasicrystal, periodically continued throughout the whole plane,
and a two-dimensional square lattice with a single step. A Penrose quasicrystal
in the axes, parallel to $\mathbf{e}_0$, $\mathbf{e}_4$ will be produced, if a
lattice is superposed on the probability distribution graph in such a way that
a point of the lattice with the coordinates $(n_0, n_4)$ should be within the
U$_{k^{\,\prime}}$ region of a corresponding structural cell (Fig.\ref{f:22})
under the number $(n_0,n_4)$.
Introduce a rotation operator $R(\varphi,r_0)$, where $\varphi$ is the rotation
angle, $r_0$ is the point coordinate, around which rotation is performed ,
then for the Penrose crystal
\begin{equation}
\label{math/62}
R(2\pi/5,r_0)\rho(\mathbf{r},\boldsymbol{\alpha},\boldsymbol{\beta})=
\rho(\mathbf{r},\boldsymbol{\alpha}^{\,\prime},\boldsymbol{\beta}^{\,\prime}).
\end{equation}
The considered quasicrystal is invariant in respect to rotations through angles,
multiple of $72^{\circ}$, i.e. it has the symmetry axis of the fifth order.
In a general case a quasicrystal will have the symmetry axis of the $q$ order,
if the following condition holds true
\begin{equation}
\label{math/63}
R(2\pi/q,r_0)\rho(\mathbf{r},\boldsymbol{\alpha},\boldsymbol{\beta})=
\rho(\mathbf{r},\boldsymbol{\alpha}^{\,\prime},\boldsymbol{\beta}^{\,\prime})
\end{equation}
In such a quasicrystal the position of lines is assigned by formula
(\ref{math/30}) with
$\chi_0=\chi_1=\ldots=\chi_k$, $\sigma_0=\sigma_1=\ldots=\sigma_k$,
$\mathbf{a}_n=a \bigl( \cos(2\pi n/q),\:\sin(2\pi n/q) \bigr)$.
The transformations (\ref{math/63}) are a cyclic group of $C_q$ rotations. A
complete list of groups of proper rotations of a two-dimensional quasicrystal
will be supplemented with a $D_q^{\,\prime}$ group of the same rotations taken
together with reflections in respect to $q$ axes. Three dimensional quasicrystals
can also be invariant in respect to the T, W, P transformation groups, retaining
invariant a regular tetrahedron, a cube (or octahedron) and a dodecahedron [9]
(or an icosahedron), respectively.
\section                       {Conclusion.}
To describe quasicrystals, in contrast to periodic crystals, use is made of a
concept on qyasicrystals as objects consisting of several structural units
having a quasiperiodic ordering. It is due to a quasiperiodic ordering of
structural units that symmetries, different from classical, are permissible for
quasicrystals. The developed approach to the description of quasicrystals
allows simple analytic expressions to be obtained for the main structural
characteristics---probabilities, which determine the repetition frequency of
corresponding structural cells. A periodic crystal can be presented as a
particular case of a quasicrystal with one structural cell and a corresponding
probability with the period 1. Such a technique of describing quasicrystals
generalizes a definition of the crystal, but  does not  complete it, since there
exists a mathematical model of aperiodic crystals\footnote{to be published}.
It is not excluded that there may exist still more complex
crystal structures.

\appendix
\section        {Function product calculation \\ of the form
$f(x,\delta_1,\delta_2)\cdot f(x-\varepsilon, \delta_3, \delta_4)$}\label{cfpr}
Function products of the form $f(x,\delta_1,\delta_2)\cdot f(x-\varepsilon,\delta_3,\delta_4)$
are calculated graphically. For generality of results consider the
$f(x,\delta_{k},\delta_{k+1})$ functions whose impulse length is greater than
the distances between the impulses $\delta_{k}-\delta_{k+1}>\sigma -\delta_{k}+\delta_{k+1}$.
\begin{figure}[ptb]
\centering
\includegraphics{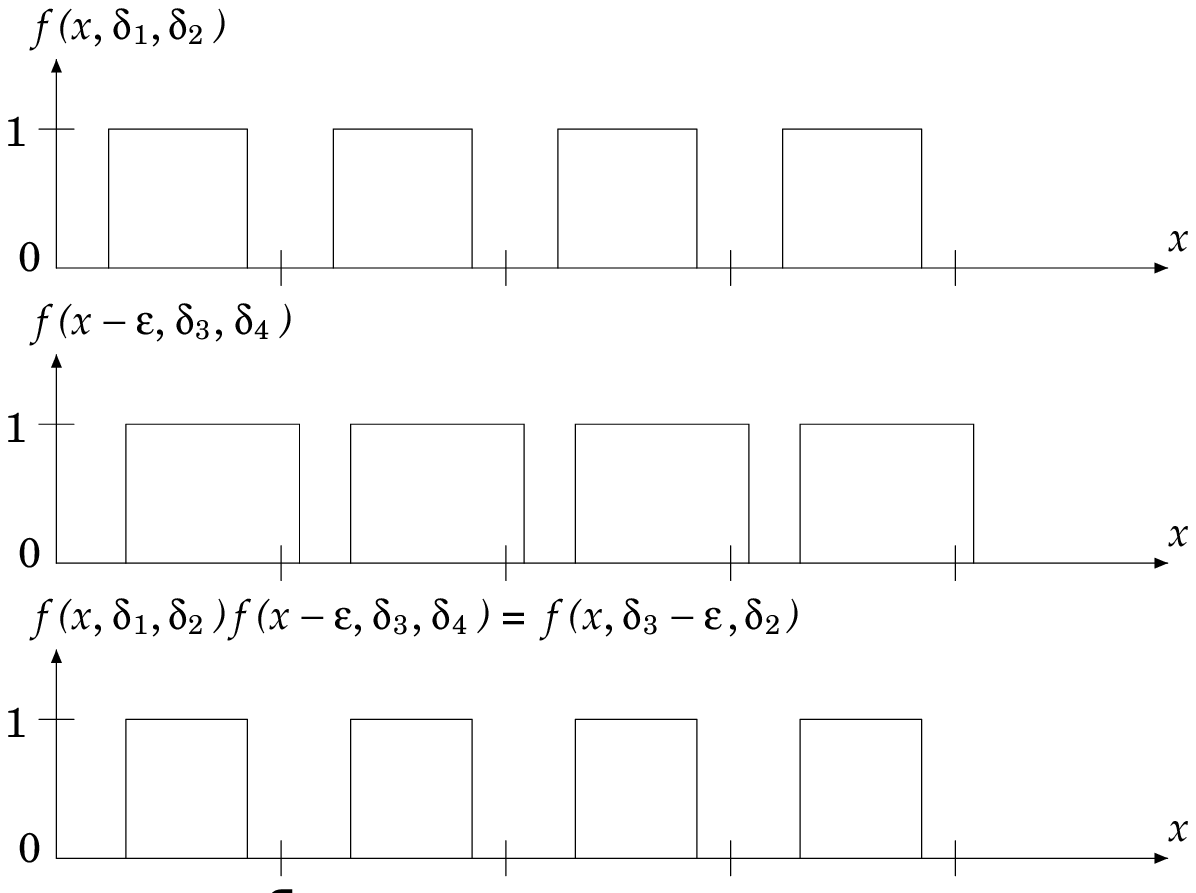}
\captn{$\varepsilon< \sigma -\delta_1+\delta_4.$}
\label{f:23}
\end{figure}
\begin{figure}[ptb]
\centering
\includegraphics*{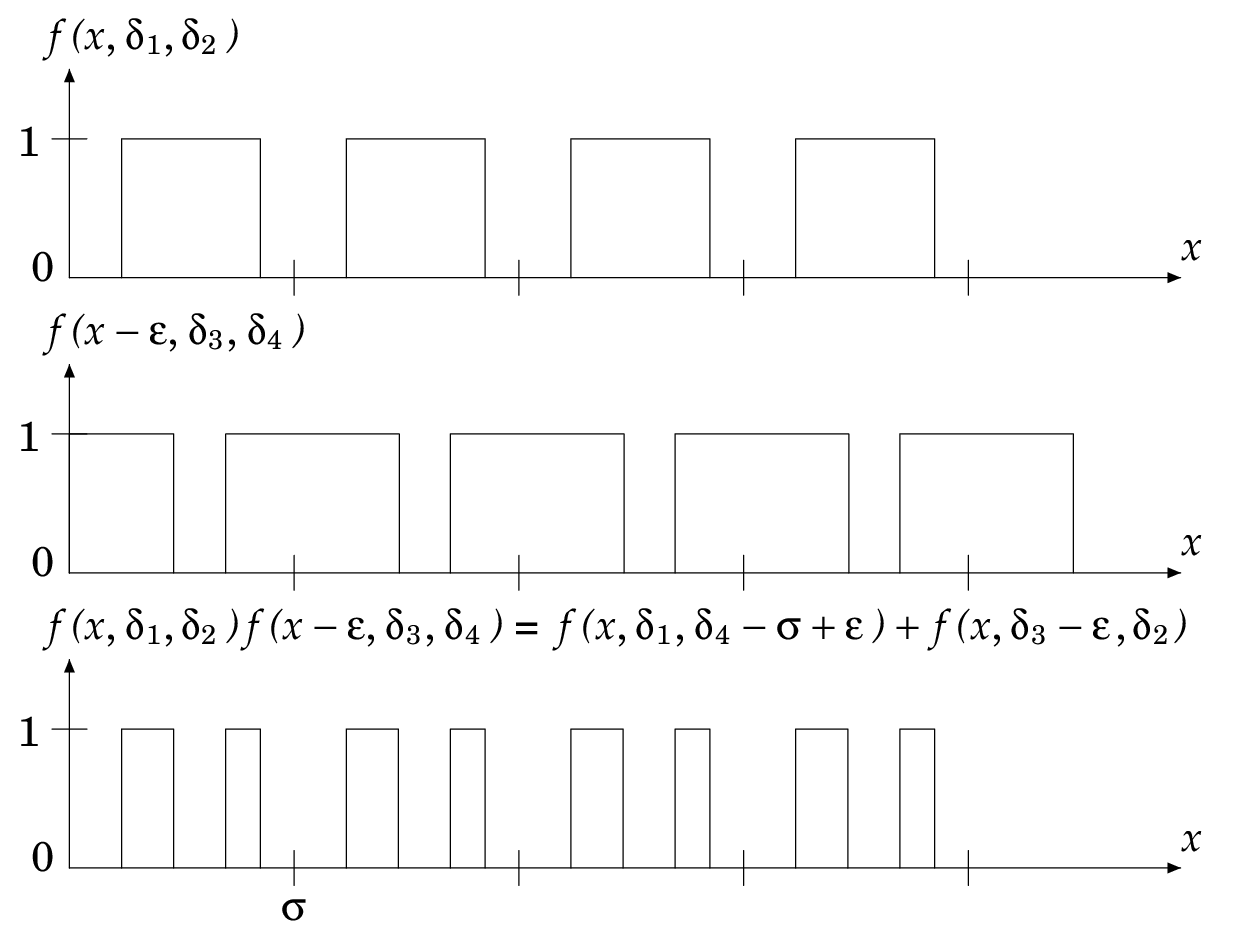}
\captn{$\sigma >\varepsilon> \sigma -\delta_1+\delta_2.$}
\label{f:24}
\end{figure} \\
(a) In the case (Fig.\ref{f:23}), when $\varepsilon< \sigma -\delta_1+\delta_4$:
\begin{equation}
\label{math/64}
f(x,\delta_1,\delta_2)f(x-\varepsilon,\delta_3,\delta_4)=
f(x,\delta_3-\varepsilon,\delta_2)
\end{equation}
when (Fig.\ref{f:24}) $\sigma >\varepsilon> \sigma -\delta_1+\delta_2$:
\begin{equation}
\label{math/65}
f(x,\delta_1,\delta_2)f(x-\varepsilon,\delta_3,\delta_4)=
f(x,\delta_1,\delta_4+\sigma-\varepsilon) +f(x,\delta_3-\varepsilon,\delta_2)
\end{equation}
(b) If $\varepsilon>\sigma$, $f(x,\delta_3,\delta_4)$ is shifted more
than by one period. Make the following transformation:
\begin{multline}
\label{math/66}
f(x-\varepsilon,\delta_3,\delta_4)= f(x-\sigma\{\varepsilon/\sigma\}-
\sigma \lfloor \varepsilon/\sigma \rfloor, \delta_3,\delta_4)= \\
f(x-\sigma\{\varepsilon/\sigma\}, \delta_3,\delta_4),
\end{multline}
after that one can consider the case (a) and (b) for a shift by the value
$\sigma \{ \varepsilon/\sigma \}$.

\section         {Analytic presentation of probabilities\\
 for a quasicrystal defined of formula \ref{math/31}}
$\beta_1 \neq \beta_2$.
\begin{subequations}
\label{math/67}
\allowdisplaybreaks
\begin{align}
p_1(\mathbf{n}+\boldsymbol{\beta})=1,& \quad \mbox{if }
     \; \left|\begin{array}{l}
     \Delta(n_0+\beta_0) \geqslant 1/\tau,\; \Delta(n_1+\beta_1) < \tau ,\\
     \Delta(n_1+\beta_1) > \Delta(n_0+\beta_0)+1/\tau ,
    \end{array}\right. \vphantom{ \cfrac{1}{ 1+\frac{1}{ 1+\frac{1}{1} } } }
\\
p_2(\mathbf{n}+\boldsymbol{\beta})=1,& \quad \mbox{if }
    \; \left| \begin{array}{l}
 \Delta(n_0+\beta_0)\geqslant 1/\tau ,\; \Delta(n_1+\beta_1)< \tau ,\\
 \Delta(n_1+\beta_1) < \Delta(n_0+\beta_0)+1/\tau,\\
 \Delta(n_1+\beta_1)> \Delta(n_0+\beta_0)/\tau+1/\tau ,\; \\
 \Delta(n_1+\beta_1) >\tau \Delta(n_0+\beta_0)-1/\tau^2 ,
    \end{array}\right. \vphantom{ \cfrac{1}{ 1+\cfrac{1}{ 1+\frac{1}{1} } } }
\\
p_3(\mathbf{n}+\boldsymbol{\beta})=1,& \quad \mbox{if }
    \; \left| \begin{array}{l}
 \Delta(n_0+\beta_0)\geqslant 1/\tau ,\\
 \Delta(n_1+\beta_1) < \Delta(n_0+\beta_0)/\tau + 1/\tau, \\
 \Delta(n_1+\beta_1)> \tau \Delta(n_0+\beta_0)-1/\tau^2 ,
     \end{array}\right. \vphantom{ \cfrac{1}{ 1+\cfrac{1}{ 1+\frac{1}{1} } } }
\\
p_4(\mathbf{n}+\boldsymbol{\beta})=1,& \quad \mbox{if }
    \; \left| \begin{array}{l}
 \Delta(n_1+\beta_1)< \tau ,  \\
 \Delta(n_1+\beta_1) > \Delta(n_0+\beta_0)/\tau - 1/\tau , \\
 \Delta(n_1+\beta_1)< \tau \Delta(n_0+\beta_0)-1/\tau^2 ,
    \end{array}\right. \vphantom{ \cfrac{1}{ 1+\cfrac{1}{ 1+\frac{1}{1} } } }
\\
p_5(\mathbf{n}+\boldsymbol{\beta})=1,& \quad \mbox{if }
    \; \left| \begin{array}{l}
 \Delta(n_1+\beta_1)> \Delta(n_0+\beta_0) ,  \\
 \Delta(n_1+\beta_1) < \Delta(n_0+\beta_0)/\tau - 1/\tau , \\
 \Delta(n_1+\beta_1)< \tau \Delta(n_0+\beta_0)-1/\tau^2 ,
    \end{array}\right. \vphantom{ \cfrac{1}{ 1+\cfrac{1}{ 1+\frac{1}{1} } } }
\\
p_6(\mathbf{n}+\boldsymbol{\beta})=1,& \quad \mbox{if }
    \; \left| \begin{array}{l}
 \Delta(n_0+\beta_0)<\tau ,\; \Delta(n_1+\beta_1) \geqslant  1/\tau ,\\
 \Delta(n_1+\beta_1)< \Delta(n_0+\beta_0)-1/\tau ,
    \end{array}\right. \vphantom{ \cfrac{1}{ 1+\cfrac{1}{ 1+\frac{1}{1} } } }
\\
p_7(\mathbf{n}+\boldsymbol{\beta})=1,& \quad \mbox{if }
    \; \left| \begin{array}{l}
 \Delta(n_0+\beta_0)< \tau ,\; \Delta(n_1+\beta_1) \geqslant  1/\tau, \\
 \Delta(n_1+\beta_1)> \Delta(n_0+\beta_0)-1/\tau ,\\
 \Delta(n_1+\beta_1)< \Delta(n_0+\beta_0)/\tau + 1/\tau^3 ,\\
 \Delta(n_1+\beta_1)> \tau \Delta(n_0+\beta_0)-1 ,
    \end{array}\right. \vphantom{ \cfrac{1}{ 1+\cfrac{1}{ 1+\frac{1}{1} } } }
\\
p_8(\mathbf{n}+\boldsymbol{\beta})=1,& \quad \mbox{if }
    \; \left| \begin{array}{l}
 \Delta(n_1+\beta_1)\geqslant 1/\tau ,\\
 \Delta(n_1+\beta_1)< \Delta(n_0+\beta_0)/\tau+1/\tau^3 ,\\
 \Delta(n_1+\beta_1)> \tau \Delta(n_0+\beta_0)-1 ,
    \end{array}\right. \vphantom{ \cfrac{1}{ 1+\cfrac{1}{ 1+\frac{1}{1} } } }
\\
p_9(\mathbf{n}+\boldsymbol{\beta})=1,& \quad \mbox{if }
    \; \left| \begin{array}{l}
 \Delta(n_0+\beta_0) < \tau ,\\
 \Delta(n_1+\beta_1) > \Delta(n_0+\beta_0)/\tau+1/\tau^3 ,\\
 \Delta(n_1+\beta_1)< \tau \Delta(n_0+\beta_0)-1 ,
    \end{array}\right. \vphantom{ \cfrac{1}{ 1+\cfrac{1}{ 1+\frac{1}{1} } } }
\\
p_{10}(\mathbf{n}+\boldsymbol{\beta})=1,& \quad \mbox{if }
    \; \left| \begin{array}{l}
 \Delta(n_0+\beta_0) > \Delta(n_1+\beta_1) ,\\
 \Delta(n_1+\beta_1) > \Delta(n_0+\beta_0)/\tau+1/\tau^3 ,\\
 \Delta(n_1+\beta_1)> \tau \Delta(n_0+\beta_0)-1 ,
    \end{array}\right. \vphantom{ \cfrac{1}{ 1+\cfrac{1}{ 1+\frac{1}{1} } } }
\\
p_{11}(\mathbf{n}+\boldsymbol{\beta})=1,& \quad \mbox{if }
    \; \left| \begin{array}{l}
 \Delta(n_0+\beta_0) \geqslant 0 ,\; \Delta(n_1+\beta_1)<\tau\\
 \Delta(n_1+\beta_1) > \Delta(n_0+\beta_0)/\tau+2/\tau ,
    \end{array}\right. \vphantom{ \cfrac{1}{ 1+\cfrac{1}{ 1+\frac{1}{1} } } }
\\
p_{12}(\mathbf{n}+\boldsymbol{\beta})=1,& \quad \mbox{if }
    \; \left| \begin{array}{l}
 \Delta(n_0+\beta_0) \geqslant 0 ,\;\\
 \Delta(n_1+\beta_1) < \Delta(n_0+\beta_0)/\tau + 2/\tau \\
 \Delta(n_1+\beta_1) > \Delta(n_0+\beta_0)+1 ,
    \end{array}\right. \vphantom{ \cfrac{1}{ 1+\cfrac{1}{ 1+\frac{1}{1} } } }
\\
p_{13}(\mathbf{n}+\boldsymbol{\beta})=1,& \quad \mbox{if }
    \; \left| \begin{array}{l}
 \Delta(n_0+\beta_0) \geqslant 0 ,\;\\
 \Delta(n_1+\beta_1) > \tau \Delta(n_0+\beta_0) + 1/\tau \\
 \Delta(n_1+\beta_1) < \Delta(n_0+\beta_0)+1 ,
    \end{array}\right. \vphantom{ \cfrac{1}{ 1+\cfrac{1}{ 1+\frac{1}{1} } } }
\\
p_{14}(\mathbf{n}+\boldsymbol{\beta})=1,& \quad \mbox{if }
    \; \left| \begin{array}{l}
 \Delta(n_0+\beta_0) < \tau ,\;\\
 \Delta(n_1+\beta_1) < \tau \Delta(n_0+\beta_0) + 1/\tau \\
 \Delta(n_1+\beta_1) > \Delta(n_0+\beta_0)+1/\tau ,
    \end{array}\right. \vphantom{ \cfrac{1}{ 1+\cfrac{1}{ 1+\frac{1}{1} } } }
\\
p_{15}(\mathbf{n}+\boldsymbol{\beta})=1,& \quad \mbox{if }
    \; \left| \begin{array}{l}
 \Delta(n_0+\beta_0) \geqslant \tau ,\\
 \Delta(n_1+\beta_1) < \Delta(n_0+\beta_0) + 1/\tau , \\
 \Delta(n_1+\beta_1) > \Delta(n_0+\beta_0)/\tau+1/\tau ,
    \end{array}\right. \vphantom{ \cfrac{1}{ 1+\cfrac{1}{ 1+\frac{1}{1} } } }
\\
p_{16}(\mathbf{n}+\boldsymbol{\beta})=1,& \quad \mbox{if }
    \; \left| \begin{array}{l}
 \Delta(n_0+\beta_0) \geqslant 1/\tau ,\; \Delta(n_1+\beta_1) \geqslant 1/\tau ,\\
 \Delta(n_1+\beta_1) < \Delta(n_0+\beta_0)/\tau+1/\tau ,
    \end{array}\right. \vphantom{ \cfrac{1}{ 1+\cfrac{1}{ 1+\frac{1}{1} } } }
\\
p_{17}(\mathbf{n}+\boldsymbol{\beta})=1,& \quad \mbox{if }
    \; \left| \begin{array}{l}
 \Delta(n_1+\beta_1) < 1/\tau ,\; \Delta(n_0+\beta_0) \geqslant \tau ,\\
 \Delta(n_1+\beta_1) > \tau \Delta(n_0+\beta_0)-1 ,
    \end{array}\right. \vphantom{ \cfrac{1}{ 1+\cfrac{1}{ 1+\frac{1}{1} } } }
\\
p_{18}(\mathbf{n}+\boldsymbol{\beta})=1,& \quad \mbox{if }
    \; \left| \begin{array}{l}
 \Delta(n_1+\beta_1) < 1/\tau ,\\
 \Delta(n_1+\beta_1) > \Delta(n_0+\beta_0) -1/\tau ,\\
 \Delta(n_1+\beta_1) < \tau \Delta(n_0+\beta_0)-1 ,
    \end{array}\right. \vphantom{ \cfrac{1}{ 1+\cfrac{1}{ 1+\frac{1}{1} } } }
\\
p_{19}(\mathbf{n}+\boldsymbol{\beta})=1,& \quad \mbox{if }
    \; \left| \begin{array}{l}
 \Delta(n_1+\beta_1) < 1/\tau ,\\
 \Delta(n_1+\beta_1) < \Delta(n_0+\beta_0) -1/\tau ,\\
 \Delta(n_1+\beta_1) < \Delta(n_0+\beta_0)/\tau-1/\tau^2 ,
    \end{array}\right. \vphantom{ \cfrac{1}{ 1+\cfrac{1}{ 1+\frac{1}{1} } } }
\\
p_{20}(\mathbf{n}+\boldsymbol{\beta})=1,& \quad \mbox{if }
    \; \left| \begin{array}{l}
 \Delta(n_1+\beta_1) \geqslant 0 ,\\
 \Delta(n_1+\beta_1) > \Delta(n_0+\beta_0) -1 ,\\
 \Delta(n_1+\beta_1) < \Delta(n_0+\beta_0)/\tau-1/\tau^2 ,
    \end{array}\right. \vphantom{ \cfrac{1}{ 1+\cfrac{1}{ 1+\frac{1}{1} } } }
\\
p_{21}(\mathbf{n}+\boldsymbol{\beta})=1,& \quad \mbox{if }
    \; \left| \begin{array}{l}
 \Delta(n_1+\beta_1) \geqslant 0 ,\\
 \Delta(n_1+\beta_1) < \Delta(n_0+\beta_0) -1 ,\\
 \Delta(n_1+\beta_1) > \tau \Delta(n_0+\beta_0)-2 ,
    \end{array}\right. \vphantom{ \cfrac{1}{ 1+\cfrac{1}{ 1+\frac{1}{1} } } }
\\
p_{22}(\mathbf{n}+\boldsymbol{\beta})=1,& \quad \mbox{if }
    \; \left| \begin{array}{l}
 \Delta(n_1+\beta_1) \geqslant 0 ,\; \Delta(n_0+\beta_0) < \tau ,\\
 \Delta(n_1+\beta_1) < \tau \Delta(n_0+\beta_0)-2 ,
    \end{array}\right. \vphantom{ \cfrac{1}{ 1+\cfrac{1}{ 1+\frac{1}{1} } } }
\\
p_{23}(\mathbf{n}+\boldsymbol{\beta})=1,& \quad \mbox{if }
    \; \left| \begin{array}{l}
  0\leqslant \Delta(n_0+\beta_0) < 1/\tau ,\\
  0 \leqslant \Delta(n_1+\beta_1) < 1/\tau  .
    \end{array}\right. 
\end{align}
\end{subequations}
In other cases the probabilities p$_{k}(\mathbf{n}+\boldsymbol{\beta})=0$.
\section*{References}
\parindent 10pt
\par
\begin{enumerate}
\item[1.] Penrose R. // Bull. Inst. Math. Appl. 1974. V.10.P.226
\item[2.] Shechtman D., Blech I., Gratias D., Cahn J.W. //Phys. Rev. Lett.,
          1984, V.53. PP 1951-1953.
\item[3.] P. Kramer // Acta Crystallogr., 1982, v. A38, P. 257-264.

\item[4.] Kramer P., Neri.R //Acta Crystallogr., 1984, v. A40, P. 580.
\item[5.] P.A. Kalugin, A. Kitaev, and L. Levitov //JETP 1985, V 41, PP 119.
\item[6.] M.Duneau and A. Katz, Phys. Rev. Lett., 1985, V 54, P 2688.
\item[7.] Sachdev S., Nelson D.R. // Phys. Rev., V. B, 1984.
\item[8.] J.E.S. Socolor, P.J Steinhardt, and D. Levine. //Phys. Rev. B, 1985,
          V 32,  5547
\item[9.] D. Levine., P.J. Steinhardt // Phys. Rev. B, 1986, V 34, P 596
\item[10.] N. de Bruijn. // Ned. Akad. Weten. Proc. Ser. A, 1981, V 43, P 27;
           1981, V 43, P 39; 1981, V 43, P 53
\item[11.] F. Gahler and J. Rhyner, // J. Math. Phys. A, 1986, V19, P 267.
\item[12.] H.-C. Jeang \& P.J. Steinhardt // Phys. Rev. Lett., 1994, V 73, PP
           1943-1946.
\item[13.] A. Janner \& T. Jansen. // Phys. Rev. B, !977, V 15, PP643-658.
\item[14.] A. Janner \& T. Jansen. // Acta Crystallogr., A, 1980, V 36, PP
           399-415.
\item[15.] A.S. Fraenkel // Canad. J. Math., 1969, V 21, PP 6-27.
\item[16.] A.S. Fraenkel, M.Mushkin, and U. Tassa. // Canad. Math. Bull., 1978, V 21, PP441-446.
\item[17.] E. N. Gilbert. // Amer. Math. Monthly, 1963, V 70, PP 736-738.
\item[18.] P.S. Kireyev //Vvedenie v teoriu grupp i eye primenenie v fizike
 tverdogo tela. 1979, M "Vysshaya shkola" s.54.
\end{enumerate}
\end{document}